\documentclass[aoas,MSNbibl,nameyear,rotating,seceqn,dvips]{arximspdf}
\usepackage{dcolumn}
\usepackage{graphicx}

\doi{10.1214/13-AOAS710} 
\volume{8}
\issue{1}
\pubyear{2014}
\firstpage{89}
\lastpage{119}

\makeatletter
\newcolumntype{d}[1]{D{.}{.}{#1}}
\newcommand{\rrvert}{\vert}
\newcommand{\llvert}{\vert}
\newcommand{\erfc}{\mathop{\operatorname{erfc}}}
\newcommand{\xmin}{x_{\min}}
\newcommand{\bmin}{b_{\min}}
\newcommand{\bminhat}{\hat{b}_{\min}}
\newcommand{\e}{\mathrm{e}}
\newcommand{\citetp}[1]{[\citet{#1}]}
\renewcommand{\d}{\mathrm{d}}
\newcommand{\eqref}[1]{(\ref{#1})}
\makeatother

\begin{document}
\begin{frontmatter}

\title{Power-law distributions in binned empirical data\thanksref{T1}} %
\thankstext{T1}{Supported in part by the Santa Fe Institute.}
\runtitle{Power-law distributions in binned empirical data}

\begin{aug}
\author[a]{\fnms{Yogesh} \snm{Virkar}\corref{}\thanksref{m1}\ead[label=e1]{yogesh.virkar@colorado.edu}}
\and
\author[b]{\fnms{Aaron} \snm{Clauset}\thanksref{m1,m3}\ead[label=e2]{aaron.clauset@colorado.edu}} 
\runauthor{Y.~Virkar and A.~Clauset}
\affiliation{University of Colorado, Boulder\thanksmark{m1} and
Santa Fe Institute\thanksmark{m3}}
\address[a]{Department of Computer Science\\
University of Colorado, Boulder\\
Boulder, Colorado 80309\\
USA\\
\printead{e1}}
\address[b]{Department of Computer Science\\
University of Colorado, Boulder\\
Boulder, Colorado 80309\\
and\\
BioFrontiers Institute\\
Santa Fe Institute\\
USA\\
\printead{e2}}
\end{aug}

\received{\smonth{8} \syear{2012}}
\revised{\smonth{12} \syear{2013}}

%
\begin{abstract}
Many man-made and natural phenomena, including the intensity of
earthquakes, population of cities and size of international wars, are
believed to follow power-law distributions. The accurate identification
of power-law patterns has significant consequences for correctly
understanding and modeling complex systems. However, statistical
evidence for or against the power-law hypothesis is complicated by
large fluctuations in the empirical distribution's tail, and these are
worsened when information is lost from binning the data. We adapt the
statistically principled framework for testing the power-law
hypothesis, developed by Clauset, Shalizi and Newman, to the case of
binned data. This approach includes maximum-likelihood fitting, a
hypothesis test based on the Kolmogorov--Smirnov goodness-of-fit
statistic and likelihood ratio tests for comparing against alternative
explanations. We evaluate the effectiveness of these methods on
synthetic binned data with known structure, quantify the loss of
statistical power due to binning, and apply the methods to twelve
real-world binned data sets with heavy-tailed patterns.
\end{abstract}

%

\end{frontmatter}

\section{Introduction}
\label{sec:intro}

Power-law distributions have attracted broad scientific interest
\citetp
{stumpf:porter:2012} both for their mathematical properties, which
sometimes lead to surprising consequences, and for their appearance in
a wide range of natural and man-made phenomena, spanning physics,
chemistry, biology, computer science, economics and the social
sciences \citetp{mitzenmacher04,newman:2005,sornette:2006,gabaix:2009}.

Quantities that follow a power-law distribution are sometimes said to
exhibit ``scale invariance,'' indicating that common, small events are
not qualitatively distinct from rare, large events. Identifying this
pattern in empirical data can indicate the presence of unusual
underlying or endogenous processes, for example, feedback loops,
network effects, self-organization or optimization, although not
always \citetp{reed:hughes:2002}. Knowing that a quantity does or does
not follow a power law provides important theoretical clues about the
underlying generative mechanisms we should consider. It can also
facilitate statistical extrapolations about the likelihood of very
large events \citetp{clauset:woodard:2012}.


Consider an empirical data example: the number of branches as a
function of their diameter for plant species \textit{Cryptomeria}. This
data was first analyzed by \citet{shinokazi:etal:1964} using the pipe
model theory. This theory claims that a tree is an assemblage of unit
pipes. If a pipe of diameter $d$ gives rise to two pipes of diameter
$d/2$ and each one of those gives rise to pipes of diameter $d/4$ and
so on, then the diameter sizes would be fractal and, hence, the
frequency distribution would follow a power law. \citet{west09} test
this theory for entire forests. However, since the original data were
binned, they used {ad hoc} methods leaving significant ambiguity
in their statistical results. Knowing whether the distribution of
branch diameters follows a power law or not is a test of the pipe model
theory and that whether a similar distribution holds for the sizes of
trees in a forest is a test for whether or not forests are simply
scaled up versions of individual plants. We reanalyze these data in
Section~\ref{sec:applications} using reliable tools we present in this
paper and show why other hypotheses might be worth considering.





Another example of interest is the hurricane intensity data measured as
the maximum wind speed in knots \citetp{nhc11}. Given the impact of
tropical storms on climate change, modeling the extreme storm events or
the upper tail of the frequency distribution is an important practical
problem. The end goal of such statistical analyses is to identify which
classes of underlying generative processes may have produced the
observed data. Identifying these mechanisms relies on correctly and
algorithmically separating the upper tail of the frequency distribution
from the body of the distribution and also reliably testing the
proposed tail model.



The task of deciding if some quantity does or does not plausibly follow
a power law is complicated by the existence of large fluctuations in
the empirical distribution's upper tail, precisely where we wish to
have the most accuracy. These fluctuations are amplified (as
information is lost) when the empirical data are binned, that is,
converted into a series of counts over a set of nonoverlapping ranges
in event size. As a result, the upper tail's true shape is often
obscured and we may be unable to distinguish a power-law pattern from
alternative heavy-tailed distributions like the stretched exponential
or the log-normal. Here, we present a set of principled statistical
methods, adapted from \citet{clauset:shalizi:newman:2009}, for
answering these questions when the data are binned.

Statistically, power-law distributions generate large events orders of
magnitude more often than would be expected under a Normal distribution
and, thus, such quantities are not well characterized by quoting a
typical or average value. For instance, the 2000 U.S. Census indicates
that the average population of a city, town or village in the United
States contains 8226 individuals, but this value gives no indication
that a significant fraction of the U.S. population lives in cities
like New York and Los Angeles, whose populations are roughly 1000 times
larger than the average. Extensive discussions of this and other
properties of power laws can be found in reviews by Mitzenmacher
(\citeyear{mitzenmacher04}), Newman (\citeyear{newman:2005}),
Sornette (\citeyear{sornette:2006}) and Gabaix (\citeyear{gabaix:2009}).

Mathematically, a quantity $x$ obeys a power law if it is drawn from a
probability distribution with a density of the form
%
\begin{equation}
p(x) \propto x^{-\alpha},
\end{equation}
where $\alpha>1$ is the \textit{exponent} or \textit{scaling parameter}
and $x>\xmin>0$. The power-law pattern holds only above some value
$\xmin$, and we say that the \textit{tail} of the distribution follows
a power law. Some researchers represent this by the use of slowly
varying functions often denoted by $L(x)$ such that the tail of the
probability density follows a power law:
%
\begin{equation}
p(x) \propto L(x) x^{-\alpha},
\end{equation}
where in the limit of large $x$, $L(cx)/L(x) \to1$ for any $c > 0$. In
testing this model with empirical data, a critical step is to determine
the cutoff or the point $\xmin$ above which the $x^{-\alpha}$ term
starts to dominate the above functional form (see Section~\ref{sec:svfun} for details).

Recently, Clauset, Shalizi and Newman \citetp
{clauset:shalizi:newman:2009} introduced a set of statistically
principled methods for fitting and testing the power-law hypothesis for
continuous or discrete-valued data. Their approach combines
maximum-likelihood techniques for fitting a power-law model to the
distribution's upper tail, a distance-based method \citetp
{reiss:thomas:2007} for automatically identifying the point $\xmin$
above which the power-law behavior holds \citetp{drees:kaufmann:1998},
a goodness-of-fit test based on the Kolmogorov--Smirnov (KS) statistic
for characterizing the fitted model's statistical plausibility, and a
likelihood ratio test \citetp{vuong:1989} for comparing it to
alternative heavy-tailed distributions.

Here, we adapt these methods to the less common but important case of
binned empirical data, that is, when we see only the frequency of
events within a set of nonoverlapping ranges. Our goal is not to
provide an exhaustive evaluation of all possible principled approaches
to considering power-law distributions in binned empirical data, but
rather the more narrow aim of adapting the popular framework of \citet
{clauset:shalizi:newman:2009} to binned data. We also aim to quantify
the impact of binning on statistical accuracy of the methods. Binning
of data often occurs when direct measurements are impractical or
impossible and only the order of magnitude is known, or when we recover
measurements from an existing histogram. Sometimes, when the original
measurements are unavailable, this is simply the form of the data we
receive and, despite the loss of information due to binning, we would
still like to make strong statistical inferences about power-law
distributions. This requires specialized tools not currently available.

Toward this end, we present maximum-likelihood techniques for fitting
the power-law model to binned data, for identifying the smallest bin
$\bmin$ for which the power-law behavior holds, for testing its
statistical plausibility, and for comparing it with alternative
distributions.\setcounter{footnote}{1}\footnote{Our code is available at
\url{http://www.santafe.edu/\textasciitilde aaronc/powerlaws/bins/}.}
These methods make no assumptions about the type of binning scheme used
and can thus be applied to linear, logarithmic or arbitrary bins. In
Sections~\ref{sec:plfit}, \ref{sec:pvalue} and \ref{sec:alternatives}
we evaluate the effectiveness of our techniques on synthetic data with
known structure, showing that they are highly accurate when given a
sample of sufficient size. Their effectiveness does depend on the
amount of information lost due to binning, and we quantify this loss of
accuracy and statistical power in several ways in Section~\ref{sec:infoloss}.

Following \citet{clauset:shalizi:newman:2009}, we advocate the
following approach for investigating the power-law hypothesis in binned
empirical data:
\begin{longlist}[1.]
\item[1.]\textit{Fit the power law.} Section~\ref{sec:plfit}. Estimate the
parameters $\bmin$ and $\alpha$ of the power-law model. (Our aim is to
model the tail of the empirical distribution which starts from the bin
$\bmin$.)
\item[2.]\textit{Test the power law's plausibility.} Section~\ref{sec:pvalue}. Conduct a hypothesis test for the fitted model. If $p\geq
0.1$, the power law is considered a plausible statistical hypothesis
for the data; otherwise, it is rejected.
\item[3.]\textit{Compare against alternative distributions.} Section~\ref{sec:alternatives}. Compare the power law to alternative heavy-tailed
distributions via a likelihood ratio test. For each alternative, if the
log-likelihood ratio is significantly away from zero, then its sign
indicates whether or not the alternative is favored over the power-law
model. \label{step:3}
\end{longlist}
The model comparison step could be replaced with another statistically
principled approach for model comparison, for example, fully Bayesian,
cross-validation or minimum description length. We do not describe
these techniques here.


Practicing what we advocate, we then apply our methods to 12 real-world
data sets, all of which exhibit heavy-tailed, possibly power-law
behavior. Many of these data sets were obtained in their binned form.
We also include a few examples from \citet{clauset:shalizi:newman:2009}
to demonstrate consistency with unbinned results. Finally, to highlight
the concordance of our binned methods with the continuous or
discrete-valued methods of \citet{clauset:shalizi:newman:2009}, we
organize our paper in a similar way. We present our work in a more
expository fashion than is necessary in order to make the material more
accessible. Proofs and detailed derivations are located in the
appendices or in the indicated references.

\section{Binned power-law distributions}
\label{sec:binPL}

Conventionally, a power-law distributed quantity can be either
continuous or discrete. For continuous values, the probability density
function (p.d.f.) of a power law is defined as
%
\begin{equation}
p(x) = C x^{-\alpha} \label{eq:PLconprob}
\end{equation}
for $x > \xmin> 0$, where $C$ is the normalization constant. For
discrete values, the probability mass function is defined in the same
way, for $x > x_{\min} > 0$, but where $x$ is an integer.

Because formulae for continuous distributions, like equation \eqref
{eq:PLconprob}, tend to be simpler than those for discrete
distributions, which often involve special functions, in the remainder
of the paper we present analysis only of the continuous case. The
methods, however, are entirely general and can easily be adapted to the
discrete case.

A binned data set is a sequence of counts of observations over a set of
nonoverlapping ranges. Let $\{ x_1, \ldots, x_N \}$ denote our $N$
original empirical observations. After binning, we discard these
observations and retain only the given ranges or bin boundaries $B$ and
the counts within them $H$. Letting $k$ be the number of bins, the bin
boundaries $B$ are denoted
%
\begin{equation}
\label{eq:binbound} B = (b_1, b_2, \ldots, b_k ),
\end{equation}
where $b_{1}>0$, $k>1$, the $i$th bin covers the interval $x \in
[b_i, b_{i+1} )$, and by convention we assume the $k$th bin extends to $+\infty$. The bin counts $H$ are denoted
%
\begin{equation}
\label{eq:bincount} H = (h_1, h_2,\ldots,h_k ),
\end{equation}
where $h_{i} = \#\{b_{i}\leq x < b_{i+1}\}$ counts the number of raw
observations in the $i$th bin.

The probability that some observation falls within the $i$th bin
is the fraction of total density in the corresponding interval:
%
\begin{eqnarray}\label{eq:PLbinnedprob}
\Pr({b_i} \leq x<{b_{i+1}}) &= &\int_{b_i}^{b_{i+1}}
p(x) \,\d x
\nonumber
\\[-8pt]
\\[-8pt]
\nonumber
&=& \frac{C}{\alpha-1} \bigl[{b_i}^{1 - \alpha} -
{b_{i+1}}^{1 -
\alpha
} \bigr].
\end{eqnarray}
Subsequently, we assume that the binning scheme $B$ is fixed by an
external source, as otherwise we would have access to the raw data and
we could apply direct methods to test the power-law hypothesis \citetp
{clauset:shalizi:newman:2009}.

To test the power-law hypothesis using binned data, we must first
estimate the scaling exponent $\alpha$, which requires choosing the
smallest bin for which the power law holds, which must be a member of
the sequence $B$, that is, it must be a bin boundary. We denote this
choice $\bmin$ in order to distinguish it from $\xmin$.

\section{Fitting power laws to binned empirical data}
\label{sec:plfit}


Many studies of empirical distributions and power laws use poor
statistical methods for this task. The most common approach is to first
tabulate the histogram and then fit a regression line to the
log-frequencies. Taking the logarithm of both sides of equation \eqref
{eq:PLconprob}, we see that the power-law distribution obeys the
relation $\ln p(x) = \ln C - \alpha\ln x$, implying that it follows a
straight line on a doubly logarithmic plot. Fitting such a straight
line may seem like a reasonable approach to estimate the scaling
parameter $\alpha$, perhaps especially in the case of binned data where
binning will tend to smooth out some of the sampling fluctuations in
the upper tail. Indeed, this procedure has a long history, being used
by Pareto in the analysis of wealth distributions in the late 19th
century \citetp{arnold:1983}, by Richardson in analyzing the size of
wars in the early 20th century \citetp{richardson:1960}, and by many
researchers since.


This na\"ive form of linear regression generates significant errors
under relatively common conditions and gives no warning of its
mistakes, and its results should not be trusted [see \citet
{clauset:shalizi:newman:2009} for a detailed explanation]. In this
section we describe a generally accurate method for fitting a power-law
distribution to binned data, based on maximum likelihood. Using
synthetically generated binned data, we illustrate its accuracy and the
inaccuracy of the na\"ive linear regression approach.

\subsection{Estimating the scaling parameter}
\label{sec:findalpha}
First, we consider the task of estimating the scaling parameter $\alpha
$. Correctly estimating $\alpha$ requires a good choice for the lower
bound $\bmin$, but for now we will assume that this value is known. In
cases where it is not known, we may estimate it using the methods given
in Section~\ref{sec:findbmin}.

The chosen method for fitting parameterized models to empirical data is
the \textit{method of maximum likelihood} which provably gives accurate
parameter estimates in the limit of large sample size \citetp
{barndorffnielsen:cox:1995,wasserman:2003,beirlant:teugels:1989,hall:1982}.
In particular, the maximum likelihood estimate $\hat{\theta}$ for the
binned or multinomial data is consistent, that is, as the sample size
$n \to\infty$, the estimate converges on the truth \citetp{rao:1957}.
Details are given in the supplementary material [\citet{suppl}] (Section~1.1). In this
section, we focus on the resulting formula's use. Here and elsewhere,
we use ``hatted'' symbols to denote estimates derived from data;
hatless symbols denote the true values, which are typically unknown.

Assuming that our observations are drawn from a power-law distribution
above the value $\bmin$, the log-likelihood function is
%
\begin{equation}
\label{eq:logL} \mathcal{L}  = n(\alpha-1) \ln\bmin+ \sum
_{i=\min}^{k} h_i \ln \bigl[{b_i}^{(1 - \alpha)}
- {b_{i+1}}^{(1 - \alpha)} \bigr],
\end{equation}
where $n=\sum_{i=\min}^{k}h_{i}$ is the number of observations in the
bins at or above $\bmin$. (We reserve $N$ for the total sample size,
that is, $N = \sum_{i=1}^{k} h_i$.) For most binning schemes, including
linearly-spaced bins, a closed-form solution for the maximum likelihood
estimator (MLE) will not exist, and the choice of $\hat{\alpha}$ must
be made by numerically maximizing equation \eqref{eq:logL} over
$\alpha
$.\footnote{Although our results are based on using the nonparametric
Nelder--Mead optimization technique, any other standard optimization
techniques such as the expectation-maximization (EM) algorithm \citetp
{mclachlan:jones:1988, cadez:etal:2002} could be used and should
produce similar results. However, using the EM technique is more
complicated than the methods we present in this section.}

When the binning scheme is logarithmic, that is, when bin boundaries
are successive powers of some constant $c$, an analytic expression for
$\hat{\alpha}$ may be obtained. Letting the bin boundaries be
$B=
(c^{s}, c^{s+1}, \ldots,c^{s+k} )$, where $s$ is the power of the
smallest bin (often 0), the MLE for $\hat{\alpha}$ is
%
\begin{equation}
\label{eq:estallog} \hat{\alpha} = 1 + {\log_{c} \biggl[1 +
\frac{1}{(s-1) - \log
_{c}\bmin
+ ({1}/{n})\sum_{i=\min}^{k} i h_i } \biggr]}.
\end{equation}
This estimate is conditional on the choice of $\bmin$ and is hence
equivalent to the well-known Hill estimator \citetp{hill:1975}. The
standard error associated with $\hat{\alpha}$ is
%
\begin{equation}
\label{eq:stddeval} \hat{\sigma} = \frac{ c^{\hat{\alpha}}-c }{ c^{(1+\hat{\alpha
})/2}\ln
c \sqrt{n} }.
\end{equation}
(Note: this expression becomes positively biased for very small values
of $n$, e.g., $c=2$, $n\lesssim50$. See Figure~2 of the supplementary
material [\citet{suppl}].)

The choice of the logarithmic spacing $c$ plays an important role in
equation~\eqref{eq:stddeval}; it also has a significant impact on our
ability to test any hypotheses (Section~\ref{sec:pvalue}) and
distinguish between different types of tail behavior
(Section~\ref{sec:alternatives}). Section~\ref{sec:infoloss} explores these issues
in detail.

\subsection{Performance of scaling parameter estimators}
\label{sec:perfal}

To demonstrate the accuracy of the maximum likelihood approach, we
conducted a set of numerical experiments using synthetic data drawn
from a power-law distribution, which were then binned for analysis. In
practical situations, we typically do not know a priori, as we do in
this section, that our data are truly drawn from a power-law
distribution. Our estimation methods choose the parameter of the best
fitting power-law form but will not tell us if the power law is a good
model of the data (or, more precisely, if the power law is not a
terrible model) or if it is a better model than some alternatives.
These questions are addressed in Sections~\ref{sec:pvalue} and \ref
{sec:alternatives}.

We drew $N=10^{4}$ random deviates from a continuous power-law
distribution with $\xmin=10$ and a variety of choices for $\alpha$. We
then binned these data using either a linear scheme, with $b_{i}=10i$
(constant width of ten), or a logarithmic scheme, with $c=2$ (powers of
two such that $b_i=10 \times2^{(i-1)}$). Finally, we fitted the
power-law form to the resulting bin counts using the techniques given
in Section~\ref{sec:findalpha}.
To illustrate the errors produced by standard regression methods, we
also estimated $\alpha$ using ordinary least-squares (OLS), on both the
p.d.f. and the complementary c.d.f., and weighted least-squares (WLS)
regression, in which we weight each bin in the p.d.f. by the number of
observations it contains.

\begin{figure}
\includegraphics{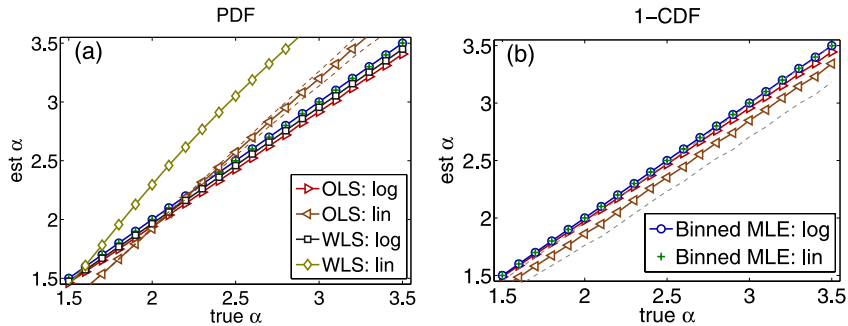}
\caption{Estimates of $\alpha$ from linearly (lin) and
logarithmically (log) binned data using maximum likelihood
and both ordinary least-squares (OLS) and weighted least-squares
(WLS) linear regression methods, using either \textup{(a)} the p.d.f. or \textup{(b)}
the complementary c.d.f. We omit error bars when they are smaller
than the symbol size. In all cases, the MLE is most accurate,
sometimes dramatically so.}
\label{fig:estalpha}
\end{figure}

Figure~\ref{fig:estalpha} shows the results, illustrating that maximum
likelihood produces highly accurate estimates, while the regression
methods that operate on doubly logarithmic plots all yield
significantly biased values, sometimes dramatically so. The especially
poor estimates for a linearly binned p.d.f. are due to the tail's very
noisy behavior: many of the upper-tail bins have counts of exactly zero
or one, which induces significant bias in both the ordinary and
weighted approaches. The regression methods yield relatively modest
bias in fitting to a logarithmically binned p.d.f. and a complementary c.d.f.
[also called a ``rank-frequency plot''---see \citet{newman:2005}],
which smooth out some of the noise in the upper tail. However, even in
these cases, maximum likelihood is more accurate.

%


Regression can be made to produce accurate estimates, using nonstandard
techniques. \citet{aban:meerschaert:2004} show that their robust
least-squares estimator has lesser variance than the $qq$-estimator of
\citet{kratz:resnick:1996} or, equivalently, the exponential tail
estimator of \citet{schultze:steinebach:1996}. Surprisingly enough,
their estimator equals the Hill estimator \citetp{hill:1975} which is
equivalent to our binned MLE. The computational complexity of this
robust least-squares linear regression approach is therefore the same
as that of our binned MLE.

\subsection{Estimating the lower bound on power-law behavior}
\label{sec:findbmin}

For most empirical quantities, the power law holds only above some
value, in the upper tail, while the body follows some other
distribution. Our goal is not to model the entire distribution, which
may have very complicated structure. Instead, we aim for the simpler
task of identifying some value $\bmin$ above which the power-law
behavior holds, estimate the scaling parameter $\alpha$ from those
data, and discard the nonpower-law data below it.


The method of choosing $\bmin$ has a strong impact on both our estimate
for $\alpha$ and the results of our subsequent tests. Choosing $\bmin$
too low may bias $\hat{\alpha}$ by including nonpower-law data in the
fit, while choosing too high throws away legitimate data and increases
our statistical uncertainty. From a practical perspective, we should
prefer to be slightly conservative, throwing away some good data if it
means avoiding bias. Unfortunately, maximum likelihood fails for
estimating the lower bound because $\bmin$ truncates the sample and the
maximum likelihood choice is always $\bmin=b_{k}$, that is, the last
bin. Some nonlikelihood-based method must be used. The common approach
of choosing $\bmin$ by visual inspection on a log--log plot of the
empirical data is obviously subjective, and thus should also be avoided.

The approach advocated in \citet{clauset:shalizi:newman:2009},
originally proposed in \citet{clauset:etal:2007}, is a distance-based
method \citetp{reiss:thomas:2007} that chooses $\xmin$ by minimizing
the distributional distance between the fitted model and the empirical
data above that choice. This approach has been shown to perform well on
both synthetic and real-world data.
Other principled approaches exist \citetp
{breiman:stone:kooperberg:1990,danielsson:etal:2001,dekkers:dehann:1993,drees:kaufmann:1998,handcock:jones:2004},
although none is universally accepted. 

Our recipe for choosing $\bmin$ is as follows:
\begin{longlist}[1.]
\item[1.]
For each possible $\bmin\in
(b_1,b_2,b_3,\ldots,b_{k-1})$, estimate $\hat{\alpha}$ using the methods
described in Section~\ref{sec:perfal} for the counts $h_{\min}$ and
higher. (For technical reasons, we require the fit to span at least two bins.)
\item[2.] Compute the Kolmogorov--Smirnov (KS) goodness-of-fit statistic\footnote{Other choices of distributional distances \citetp
{press:etal:1992} are possible options, for example, Pearson's $\chi
^{2}$ cumulative test statistic. In practice, like \citet
{clauset:shalizi:newman:2009}, we find that the KS statistic is superior.}
between the fitted c.d.f. and the empirical distribution.
\item[3.] Choose as $\bminhat$ the bin boundary with the smallest KS statistic.
\end{longlist}
The KS statistic is defined in the usual way \citetp{press:etal:1992}.
Let $P(b | \hat{\alpha},\bmin)$ be the c.d.f. for the binned power law,
with parameter $\hat{\alpha}$ and current choice $\bmin$, and let
$S(b)$ be the cumulative binned empirical distribution for counts in
bins $\bmin$ and higher. We choose $\bminhat$ as the value that minimizes
%
\begin{equation}
\label{eq:ksstat} D = \max_{ b \geq\bmin} \bigl\llvert S(b) - P(b |
\alpha,\bmin)\bigr\rrvert .
\end{equation}

Thus, when $\bmin$ is too low, reaching into the nonpower-law portion
of the empirical data, the KS distance will be high because the
power-law model is a poor fit to those data; similarly, when $\bmin$ is
too high, the sample size is small and the KS distance will also be
high. Both effects are small when $\bmin$ coincides with the beginning
of the power-law behavior.

To illustrate the accuracy of this method, we compare its performance
with the one proposed by \citet{reiss:thomas:2007}. Their methodology
(the RT method) selects the bin boundary with index $\min\in\{
1,2,3,\ldots,k-1\}$ that minimizes
%
\begin{eqnarray}
\label{eq:RT} \frac{1}{ (\sum_{j=\min}^{k} h_j )} \sum_{i=\min
}^{k}
\Biggl( \bigl\llvert \hat\alpha_{i} - \operatorname{median}(\hat\alpha_{\min},
\ldots, \hat\alpha_{k}) \bigr\rrvert { \Biggl(\sum
_{j=i}^{k} h_j \Biggr)}^{\beta
}
\Biggr),
\nonumber
\\[-8pt]
\\[-8pt]
 \eqntext{\displaystyle 0 \leq\beta< \frac{1}{2},}
\end{eqnarray}
where $(\hat\alpha_{\min}, \ldots, \hat\alpha_{k})$ are the slope
estimates calculated by considering data above bin boundaries $(b_{\min
}, \ldots, b_{k})$, respectively.

The idea behind this approach is to minimize the asymptotic mean
squared error in $\hat\alpha$ using a finite sample. The choice that
yields this minimum is the optimal sample fraction, which is the
fraction of observations after $\bmin$, and $\beta$ is a smoothing
parameter that can be used to improve the choice of $\bmin$ for small
and medium sized samples. 


\subsection{Performance of lower bound estimator}
\label{sec:perbmin}

We evaluate the accuracy of these two methods using synthetic data
drawn from a composite distribution that follows a power law above some
choice of $\bmin$ but some other distribution below it. We then apply
both linear and logarithmic binning schemes, for a variety of choices
of the true $\bmin$. The form of our test density is
%
%
\begin{equation}
p(x) = \cases{ C e^{-{\alpha}(x/\bmin- 1)}, & \quad $ b_{1} \leq x < \bmin,$\vspace
*{2pt}
\cr
C {(x/\bmin)}^{-\alpha}, & \quad$\mbox{otherwise},$} \label{eq:test:pdf}
\end{equation}
which has a continuous slope at $\bmin$ and thus departs slowly from
the power-law form below this point. This provides a difficult task for
the estimation procedure.

In our numerical experiments, we fix the sample size at $N = 10\mbox{,}000$
and use a linear scheme, $b_i = 1 + 50(i-1)$ (constant width of 50),
and a logarithmic one, $b_i = 2^{(i-1)}$ (powers of 2). For our first
experiment, we hold the scaling parameter fixed at $\alpha=3$ and
characterize each method's ability to recover the true threshold $\bmin
$, which we vary across the values of $B$. In a second experiment, we
fix $\bmin$ at the tenth bin boundary and characterize the impact of
misestimating $\bmin$ on the estimated scaling parameter, and so vary
$\alpha$ over the interval $[1.5, 3.5]$.

%
\begin{figure}
\includegraphics{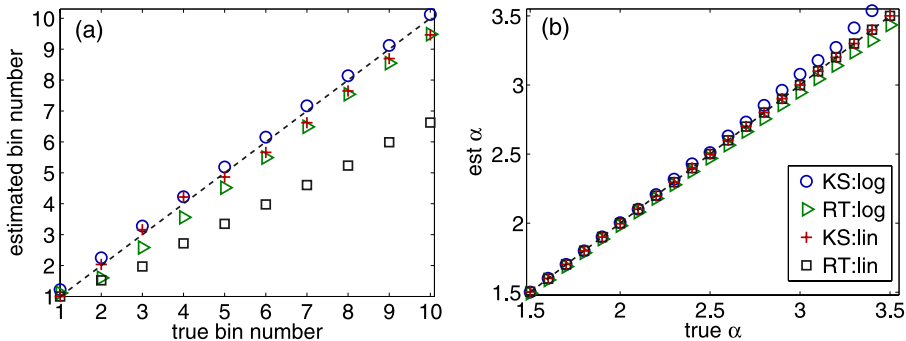}
\caption{Estimated $\bmin$ using the KS-minimization method and the
Reiss--Thomas or RT method; \textup{(a)} the true bin number versus estimated bin
number and \textup{(b)} true $\alpha$ versus $\hat{\alpha}$ for true bin number
$10$. In both figures, we show results for logarithmic (log) and linear
(lin) binning schemes with $y=x$ (dashed line) shown for reference.}
\label{fig:bin:est}
\end{figure}


Figure~\ref{fig:bin:est}(a) shows the results for estimating the
threshold, which is reliably identified in the logarithmic binning
scheme but slightly underestimated for the KS method and moderately
underestimated for the RT method with the linear scheme. However,
Figure~\ref{fig:bin:est}(b) shows that for both methods and binning
schemes, if we treat $\bmin$ as a nuisance parameter, the scaling
parameter itself is accurately estimated.

The slight deviations from the $y=x$ line in both figures highlight
some of the pitfalls of working with binned data and power-law
distributions. First, in estimating $\bmin$ [Figure~\ref{fig:bin:est}(a)],
the linear binning scheme yields a slight but
consistent underestimate, thereby including some nonpower-law data in
the estimation, while the logarithmic scheme shows no such bias. This
arises from the differences in linear versus logarithmic binning.
Because logarithmic bins span increasingly large intervals, the
distribution's curvature around $\bmin$ is accentuated, presenting a
more obvious target for the algorithm, while a linear scheme spreads
this curvature across several bins. For both algorithms the choice of
$\bminhat$ is slightly below $\bmin$, however, this does not induce a
substantial bias in $\alpha$, which remains close to the true value
[Figure~\ref{fig:bin:est}(b)].

Second, when the true value is $\alpha\gtrsim3$,\footnote{Here and
elsewhere, we use the symbols $\gtrsim$ and $\lesssim$ to mean
``approximately greater than'' and ``approximately less than,''
respectively.} we see a slight underestimation of $\alpha$ for the
RT-method caused by the slight underestimate of $\bmin$ with this
method. However, the RT-method can be shown to work in the limit of
large sample size, as this underestimation reduces with a higher $N$.
The slight overestimate of $\alpha$ for the KS-method under a
logarithmic scheme is caused by a special kind of small sample bias.
This bias appears either when the number of observations or the number
of bins in the tail region is small.

\begin{figure}
\includegraphics{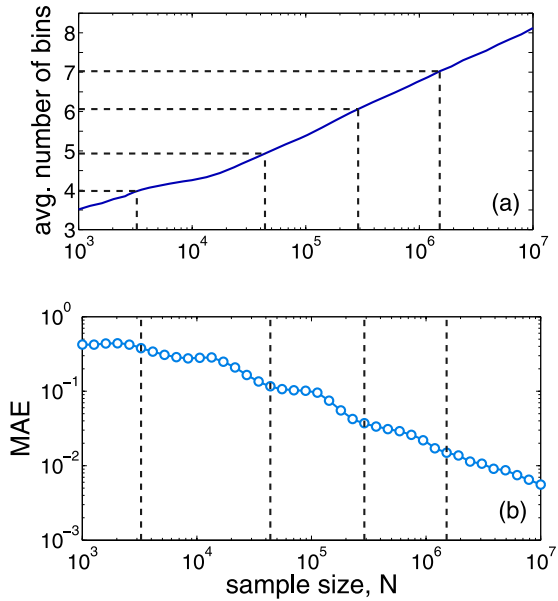}
\caption{The ``few bins'' bias; for fixed $\alpha=3.5$, $\bmin=2^{9}$
and $c=2$ logarithmic binning scheme, the \textup{(a)} average number of bins
above $\bmin$ and \textup{(b)} mean absolute error, as a function of sample size~$N$, illustrating a second-order bias that decreases as the average
number of bins in the fitted region increases.}
\label{fig:avgbinsmae}
\end{figure}

To illustrate this ``few bins'' bias, even when sample size is large,
we conduct a third experiment: using the same powers-of-two binning
scheme, we now fix $\bmin=2^{9}$ and $\alpha=3.5$, while varying the
sample size $N$. As $N$ increases, a larger number of bins above $\bmin
$ will be populated, and we measure the accuracy of $\hat{\alpha}$ as
this number increases. Figure~\ref{fig:avgbinsmae} shows that the bias
in $\hat{\alpha}$ decreases with sample size, as we expect, but with a
second-order variation that decreases as the average number of bins in
the tail region increases. The implication is that researchers must be
cognizant of both small sample issues and having too few bins in the
scaling region.

Last, obtaining uncertainty in our estimates (${\hat{b}}_{\min}$,
$\hat
\alpha$) can be done using a nonparametric bootstrap\footnote{Note that the use of bootstrap for estimating the uncertainty
is not problematic since the distributions of both $\bminhat$ and
$\hat
\alpha$ are well concentrated around their true values.}
method \citetp{efron:tibshirani:1993}.
Given empirical data with $N$ observations that are binned using a
binning scheme $B$, we generate $m$ synthetic data sets in the
following way:
\begin{longlist}[1.]
\item[1.] For each data set, we draw $N$ samples $\{x_1, \ldots, x_N \}$
such that the probability of some sample $x_j$ being drawn from the
$i$th bin is simply the cell probability $h_i/N$, where $h_i$ is the
number of observations in the $i$th bin.
%
\item[2.] We then bin the samples $\{x_1, \ldots, x_N \}$ using the same
binning style $B$.
\end{longlist}
Using these $m$ data sets, we report the standard deviation from $m$
estimates of the model parameters calculated using the methods
described above.

%


\subsection{Slowly varying functions}
\label{sec:svfun}

A function $L(x)$ is said to be slowly varying if
%
\begin{equation}
\lim_{x\to\infty} \frac{L(cx)}{L(x)} = 1
\end{equation}
for some constant $c>0$. From the perspective of extreme value theory,
we can use this notion to describe a probability density $p(x)$ that
asymptotically follows the power law as
%
\begin{equation}
p(x) \propto L(x) x^{-\alpha}.
\end{equation}
The difficulty of using this model when analyzing empirical data is the
difficulty of choosing the value of $x$ above which the $x^{-\alpha}$
term starts dominating the above equation, that is, to estimate $\xmin$
(or $\bmin$ for binned data). A common approach is to visually inspect
the plot of the estimate $\hat\alpha$ as a function of $\bmin$ (called
a Hill plot) and identify the point ${\hat{b}}_{\min}$ beyond which
$\hat\alpha$ appears stable. However, other approaches such as those of
\citet{kratz:resnick:1996} (using a $qq$-plot) and \citet
{stoev:michailidis:taqqu:2006} (using block-maxima of the data) often
yield better results.

The KS method described above can accurately estimate $\bminhat$ when
the true $\bmin$ lies in the range of the empirical data. Thus, we risk
rejecting a true power-law hypothesis, when the empirical distribution
follows the power law only for higher $x$ values not seen in the sample
or, in other words, when the $L(x)$ term dominates the $x^{-\alpha}$
term for the entire range of the sample. The general solution is to
model the structure of $L(x)$ correctly. This is highly nontrivial, as
$L(x)$ can have many parametric forms and thus testing for $L(x)$ is
difficult. The advantage of our method is that we do not model $L(x)$,
as we simply ignore the data below $\bmin$. This makes our method
inherently conservative, in that we may fail to find a power law in the
upper tail because $L(x)$ creates systematic deviations in the observed range.

Note that, in practice, quantities that follow the power law only
asymptotically may not appear to follow the power law for a measurable
sample of those quantities. Thus, an empirical sample would only be
modeled by some $L(x)$ and would not imply the interesting underlying
mechanisms that the power laws imply. Thus, if we care only about
empirical power-law distributions that can actually be measured, the
methods we describe are a reasonable approach.



\section{Testing the power-law hypothesis}
\label{sec:pvalue}
\begin{figure}[b]
\includegraphics{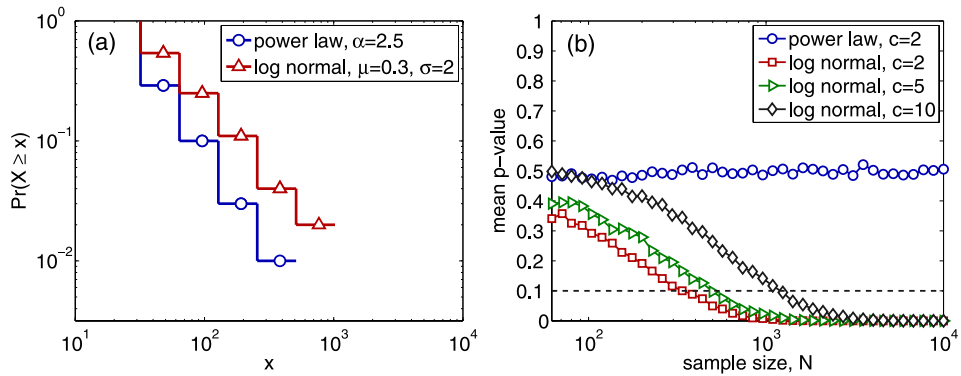}
\caption{\textup{(a)} Logarithmic histograms for two $N=100$ samples from a
power law and a log-normal distribution. Both exhibit a linear pattern
on log--log axes, despite only one being a power law. \textup{(b)}~Mean $p$ for
the power-law hypothesis, as a function of sample size $N$; dashed line
gives the threshold for rejecting the power law. For power-law data,
$p$ is typically high, while for the nonpower-law data, $p$ is a
decreasing function of sample size. Notably, the binning scheme's
coarseness determines the sample size required to correctly reject the
power-law model.}
\label{fig:pvalue}
\end{figure}

The methods of Section~\ref{sec:plfit} allow us to accurately fit a
power-law tail model to binned empirical data. These methods, however,
provide no warning if the fitted model is a poor fit to the data, that
is, when the power-law model is not a plausible generating distribution
for the observed bin counts. Because a wide variety of heavy-tailed
distributions, such as the log-normal and the stretched exponential
(also called the Weibull), among others, can produce samples that
resemble power-law distributions [see Figure~\ref{fig:pvalue}(a)], this
is a critical question to answer.

%
%

Toward this end, we adapt the goodness-of-fit test of \citet
{clauset:shalizi:newman:2009} to the context of binned data.
Demonstrating that the power-law model is plausible, however, does not
determine whether it is more plausible than alternatives. To answer
this question, we adapt the likelihood ratio test of \citet
{clauset:shalizi:newman:2009} to binned data in Section~\ref{sec:alternatives}.
For both, we additionally explore the impact of
information loss from binning on the statistical power of these tests.


\subsection{Goodness-of-fit test}
\label{sec:gof}
Given the observed bin counts and a hypothesized power-law distribution
from which the counts were drawn, we would like to know whether the
power law is plausible, given the counts.

A goodness-of-fit test provides a quantitative answer to this question
in the form of a $p$-value, which in turn represents the likelihood
that the hypothesized model would generate data with a more extreme
deviation from the hypothesis than the empirical data. If $p$ is large
(close to 1), the difference between the data and model may be
attributed to statistical fluctuations; if it is small (close to 0),
the model is rejected as an implausible generating process for the
data. From a theoretical point of view, failing to reject is sufficient
license to proceed, provisionally, with considering mechanistic models
that assume or generate a power law for the quantity of interest.

The first step of our approach is to fit the power-law model to the bin
counts, using methods described in Section~\ref{sec:plfit} to choose
$\hat{\alpha}$ and $\bminhat$. Given this hypothesized model $M$, the
remaining steps are as follows; in each case, we always use the fixed
binning scheme $B$ given to us with the empirical data:
\begin{longlist}[1.]
\item[1.]
Compute the distance $D^{*}$ between the
estimated model $M$ and the empirical bin counts $H$, using the KS
goodness-of-fit statistic, equation \eqref{eq:ksstat}.%
%
%
\item[2.]
Using a semi-parametric bootstrap, generate a
synthetic data set with $N$ values that follows a binned power-law
distribution with parameter $\hat{\alpha}$ at and above $\bminhat$, but
follows the empirical distribution below $\bminhat$. Call these
synthetic bin counts $H'$.
\item[3.]
Fit the power-law model to $H'$, yielding a
new model $M'$ with parameters $\hat{b}'_{\min}$ and $\alpha'$.
\item[4.]
Compute the distance $D$ between $M'$ and $H'$.
\item[5.]
Repeat steps 2--4
many times, and report $p = \Pr(D \geq D^{*})$, the fraction of these
distances that are at least as large at $D^{*}$.
\end{longlist}
To generate synthetic binned data, the semi-parametric bootstrap in
step 2 is as follows. Recall that $n$ counts the number
of observations from the data $H$ that fall in the power-law region.
With probability $n/N$, generate a nonbinned power-law random
deviate \citetp{clauset:shalizi:newman:2009} from $M$ and increment the
corresponding bin count in the synthetic data set; otherwise, with
probability $1-n/N$, increment the count of a bin $i$ below $\bminhat$
chosen with probability proportional to its empirical count $h_{i}$.
Repeating this process $N$ times, we generate a complete synthetic data
set with the desired properties.


The use of the KS statistic as a goodness-of-fit measure is
nontraditional since it gets underestimated for binned data \citetp
{noether:1963}. However, estimating the distribution of distances in
step 4 via Monte Carlo allows us to correctly construct
the hypothesis test and choose the critical value. [As an example, see
Table~2 of \citet{goldstein:morris:yen:2004}.] This is necessary to
produce an unbiased estimate of $p$ because our original model
parameters $M$ are estimated from the empirical data. The
semi-parametric bootstrap ensures that the subsequent values $D$ are
estimated in precisely the same way---by estimating both $\bmin$ and
$\alpha$ from the synthetic data---that we estimated $D^{*}$ from $H$.
Failure to estimate $\hat{b}'_{\min}$ from $H'$, using $\bminhat$ from
$H$ instead, yields a biased and thus unreliable $p$-value.




How many such synthetic data sets should we generate? The answer given
by \citet{clauset:shalizi:newman:2009} also holds in the case of binned
data. We should generate at least $\frac{1}{4}\varepsilon^{-2}$ synthetic
data sets to achieve an accuracy of knowing $p$ to within $\varepsilon$ of
the true value. For example, if we wish to know $p$ to within $\varepsilon
=0.01$, we should generate about 2500 synthetic data sets.

Given an estimate of $p$, we must decide if it is small enough to
reject the power-law hypothesis. We recommend the relatively
conservative choice of ruling out the power law if $p<0.1$. By not
using smaller rejection thresholds, we avoid letting through some
quantities that in fact have only a small chance of actually following
a power law.

Note that a large value of $p$ does not imply the correctness of the
power law for the data. A large $p$ can arise for at least two reasons.
First, there may be alternative distributions that fit the data as well
or better than the power law, and other tests are necessary to make
this determination (which we cover in Section~\ref{sec:alternatives}).
Second, for small values of $n$, or for a small number of bins above
$\bmin$, the empirical distribution may closely follow a power-law
shape, yielding a large $p$, even if the underlying distribution is not
a power law. This happens not because the goodness-of-fit test is
deficient, but simply because it is genuinely hard to rule out the
power law if we have very little data. For this reason, a large $p$
should be interpreted cautiously either if $n$ or the number of bins in
the fitted region is small.

Finally, we do not recommend the use of the well-known $\chi^2$
statistic for analyzing heavy-tailed distributions, since it requires
the expected cell frequencies to be above a certain threshold \citetp
{tate:hye:1973} and has less statistical power compared to the KS
statistic \citetp{horn:1977}.

\subsection{Performance of the goodness-of-fit test}
\label{sec:pergof}

To demonstrate the effectiveness of our goodness-of-fit test for binned
data, we drew various-sized synthetic data from two distributions: a
power law with $\alpha=2.5$ and a log-normal distribution with $\mu
=0.3$ and $\sigma=2.0$, both with $\bmin=16$. Using these methods, one
can compare against any other alternative distribution. However, the
choice of log-normal provides a strong test because for a wide range of
sample sizes it produces bin counts that are reasonably power-law-like
when plotted on log--log axes [Figure~\ref{fig:pvalue}(a)]. 

Figure~\ref{fig:pvalue}(b) shows the average $p$-value, as a function of
sample size $N$, for the power-law hypothesis when data are drawn from
these distributions. When we fit the correct model to the data, the
resulting $p$-value is uniformly distributed, and the mean $p$-value is
$0.5$, as expected. When applied to log-normal data, however, the
$p$-value remains above our threshold for rejection only for small
samples ($N\lesssim300$), and we correctly reject the power law for
larger samples. We note, however, that the sample size at which the
$p$-value leads to a correct rejection of the power law depends on the
binning scheme, requiring a larger sample size when the binning scheme
is more coarse (larger $c$).

\section{Alternative distributions}
\label{sec:alternatives}
The methods described in Section~\ref{sec:pvalue} provide a way to test
whether our binned data plausibly follow a power law. However, many
distributions, not all of them heavy tailed, can produce data that
appear to follow a power law when binned. A large $p$-value for the
power-law model provides no information about whether some other
distribution might be an equally plausible or even a better
explanation. Demonstrating that such alternatives are worse models of
the data strengthens the statistical argument in favor of the power law.

There are several principled approaches to comparing the power-law
model to alternatives, for example, cross-validation \citetp
{stone:1974}, minimum description length \citetp{grunwald:2007} or
Bayesian techniques \citetp{kass:raftery:1994}. Following \citet
{clauset:shalizi:newman:2009}, we construct a \textit{likelihood ratio
test} proposed by \citet{vuong:1989} (LRT) for binned data. This
approach has several attractive features, including the ability to fail
to distinguish between the power law and an alternative, for example,
due to small sample sizes. Information loss from binning reduces the
statistical power of the LRT and, thus, its results for binned data
should be interpreted cautiously. Further, although there are generally
an unlimited number of alternative models, only a few are commonly
proposed alternatives or correspond to common theoretical mechanisms.
We focus our efforts on these, although in specific applications, a
researcher must use their expert judgement as to what constitutes a
reasonable alternative.

In what follows, we will consider four alternative distributions, the
exponential, the log-normal and the stretched exponential (Weibull)
distribution, plus a power-law distribution with exponential cutoff.
Table~\ref{table:pdfs} gives the mathematical forms of these models. In
application to binned data, a piecewise integration over bins, like
equation \eqref{eq:PLbinnedprob}, was carried out and parameters were
estimated by numerically maximizing the log-likelihood function.

%
\begin{table}
\caption{Definitions of alternative distributions for our likelihood
ratio tests. For each, we give the basic functional form $f(x)$ and the
appropriate normalization constant $C$ such that $\int_{\xmin}^\infty
Cf(x)\,\d x=1$ for the continuous case}\label{table:pdfs}
\begin{tabular*}{\textwidth}{@{\extracolsep{\fill}}lcc@{}}
\hline
 &
\multicolumn{2}{c}{\textbf{Density} $\bolds{p(x)=Cf(x)}$} \\[-6pt]
&
\multicolumn{2}{c@{}}{\hrulefill} \\
\multicolumn{1}{@{}l}{{\textbf{Name}}}& $\bolds{f(x)}$ & $\bolds{C}$ \\
\hline
{Power law with cutoff}
& {$x^{-\alpha}\e^{-\lambda x}$}
& {${\lambda^{1-\alpha}\over\Gamma(1-\alpha,\lambda\xmin)}$}
\\[3pt]
{Exponential}
& {$\e^{-\lambda x}$}
& {$\lambda\e^{\lambda\xmin}$} \\[3pt]
{Stretched exponential}
& {$x^{\beta-1} \e^{-\lambda x^\beta}$}
& {$\beta\lambda\e^{\lambda\xmin^\beta}$} \\[3pt]
{Log-normal}
& {${1\over x} \exp [ -{(\ln x-\mu)^2\over2\sigma^2}  ]$}
& {$\sqrt{{2\over\pi\sigma^2}} [ \erfc ( {\ln
x_{\mathrm
{min}}-\mu\over\sqrt{2}\sigma}  )  ]^{-1}$} \\
\hline
\end{tabular*}
\end{table}

\subsection{Direct comparison of models}
\label{sec:dcompmodels}
Given a pair of parametric models $A$ and $B$ for which we may compute
the likelihood of our binned data, the model with the larger likelihood
is a better fit. The logarithm of the ratio of the two likelihoods
$\mathcal{R}$ provides a natural test statistic for making this
decision: it is positive or negative depending on which distribution is
better, and it is indistinguishable from zero in the event of a tie.

Because our empirical data are subject to statistical fluctuations, the
sign of $\mathcal{R}$ also fluctuates. Thus, its direction should not
be trusted unless we may determine that its value is probably not close
to $\mathcal{R}=0$. That is, in order to make a firm choice between
distributions, we require a log-likelihood ratio that is sufficiently
positive or negative that it could not plausibly be the result of a
chance fluctuation from zero.

The log-likelihood ratio is defined as
%
\begin{equation}
\label{eq:likeliR} {\mathcal{R}}  = \ln \biggl(\frac{{\mathcal{L}_A}(H | {\hat{\theta
}}_A)}{{\mathcal{L}_B}(H | {\hat{\theta}}_B)} \biggr),
\end{equation}
where by convention $\mathcal{L}_{A}$ is the likelihood of the model
under the power-law hypothesis, fitted using the methods in Section~\ref{sec:plfit}, and $\mathcal{L}_{B}$ is the likelihood under the
alternative distribution, again fitted by maximum likelihood. To
guarantee the comparability of the models, we further require that they
be fitted to the same bin counts, that is, to those at or above
$\bminhat$ chosen by the power-law model.\footnote{This requirement is particular to the problem of fitting tail
models, where a threshold that truncates the data must be chosen. An
interesting problem for future work is thus to determine how to compare
models with different numbers of observations, as would be the case if
we let $\bmin$ vary between the two models.}

Given $\mathcal{R}$, we use the method proposed by Vuong \citetp
{vuong:1989} to determine if the observed sign of $\mathcal{R}$ is
statistically significant. This yields a $p$-value: if $p$ is small
(say, $p<0.1$), then the observed sign is not likely due to chance
fluctuations around zero; if $p$ is large, then the sign is not
reliable and the test fails to favor one model over the other.
Technical details of the likelihood ratio test are given in the supplementary material [\citet{suppl}] (Section~2). Results from \citet
{clauset:shalizi:newman:2009} show that this hypothesis test
substantially increases the reliability of the likelihood ratio test,
yielding accurate answers for much smaller data sets than if the sign
is interpreted without regard to its statistical significance.

Before evaluating the performance of the LRT on binned data, we make a
few cautionary remarks about \textit{nested models}. When one model is
strictly a subset of the other, as in the case of a power law and a
power law with exponential cutoff, even if the smaller model is the
true model, the larger model will always yield at least as large a
likelihood. In this case, we must slightly modify the hypothesis test
for the sign of $\mathcal{R}$ and use a little more caution in
interpreting the results; see supplementary material [\citet{suppl}] (Section~2).

\subsection{Performance of the likelihood ratio test}
\label{sec:loglikelytests}

%
\begin{figure}[b]
\includegraphics{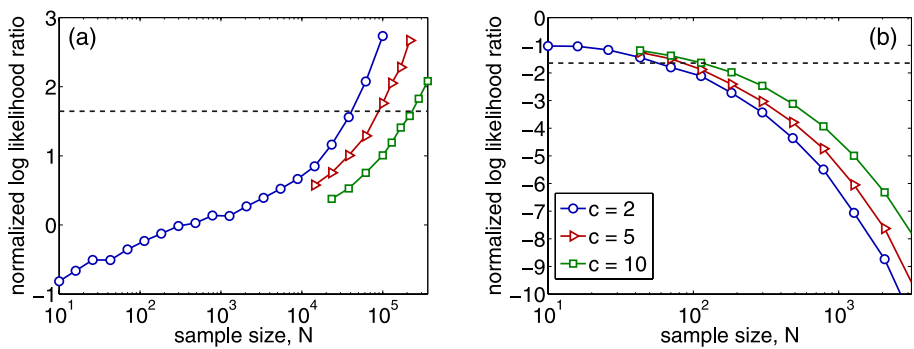}
\caption{Behavior of normalized log-likelihood
ratio $n^{-1/2} \mathcal{R}/\sigma$, for synthetic data sets drawn from
\textup{(a)} power-law and \textup{(b)} log-normal distributions. Both were then binned
using a logarithmic binning scheme, with bin boundaries in powers of
$c=\{2,5,10\}$. Dashed line indicates the threshold at which the sign
of $n^{-1/2} \mathcal{R}/\sigma$ becomes trustworthy.}
\label{fig:LRT:results}
\end{figure}
%

We demonstrate the performance of the likelihood ratio test for binned
data by pitting the power-law hypothesis against the log-normal
hypothesis. For a log-normal distribution, increasing the $\sigma$
parameter results in a power-law-like region for a large range. Thus,
in general, rejecting a log-normal hypothesis in favor of a power-law
hypothesis using any model comparison test is a difficult task, made
all the more difficult when the data are binned (see Section~\ref{sec:infoloss}). We also note that since a power law implies different
generative mechanisms as opposed to a log normal \citetp
{mitzenmacher04}, favoring one hypothesis over the other has important
scientific implications for understanding what processes generated the data.

To illustrate these points, we conduct two experiments: one in which we
draw a sample from a power-law distribution, with $\alpha= 2.5$ and
$\xmin= 1$, and a second in which we draw a sample from a log-normal
distribution, with $\mu= 0.3$ and $\sigma= 2$. We then bin these
samples logarithmically, with $c=\{2,5,10\}$, and fit and compare the
power-law and log-normal models. The normalized log-likelihood ratio
$n^{-1/2} \mathcal{R}/\sigma$ (see Section~2 of the supplementary material [\citet{suppl}])
provides a concrete measure by which to compare outcomes at different
sample sizes. If the test performs well, in the first case, $\mathcal
{R}$ will tend to be positive, correctly favoring the power law as the
better model, while in the second, the ratio will tend to be negative,
correctly rejecting the power law.

Figure~\ref{fig:LRT:results} shows the results. When the power-law
hypothesis is correct [Figure~\ref{fig:LRT:results}(a)], the sign of $\mathcal
{R}$ allows us to correctly rule in favor of the power law when the
sample size is sufficiently large. However, the size required for an
unambiguously correct decision grows with the coarseness of the binning
scheme (larger $c$). Interestingly, a reliably correct decision in
favor of the power law [Figure~\ref{fig:LRT:results}(a)] requires a much larger
sample size ($n\approx20\mbox{,}000$ here) than a decision against it
[Figure~\ref{fig:LRT:results}(b)] ($n\lesssim200$). This illustrates the
difficulty of rejecting alternative distributions like the log-normal,
which can imitate a power law over a wide range of sample sizes.

\section{Information loss due to binning}
\label{sec:infoloss}

The above results already demonstrate that binned data can make
accurately fitting and testing the power-law hypothesis more difficult.
Figure~1 of the supplementary material [\citet{suppl}] and Figures~\ref{fig:pvalue} and
\ref
{fig:LRT:results} show that all the three steps of our framework have
reduced statistical power if we use coarser binning schemes. In this
section we quantify the impact of different binning schemes on both the
statistical and model uncertainty for the power-law hypothesis.

To illustrate this loss of information, we pose the following question:
Suppose we have a sample $n_1 \to\infty$ and a binning scheme $c_1$
(logarithmic in powers of $c_1$). Given a choice of $\alpha$, how much
larger a sample do we need in order to achieve the same statistical
accuracy in $\hat\alpha$ using a coarser scheme $c_2 > c_1$?


In the limit of large sample size, the asymptotic variance is equal to
the inverse of Fisher information [see \citet{cramer46,rao46}]. For the
two different binning schemes ($c_1$, $c_2$) and the corresponding
sample sizes ($n_1, n_2$), the following approximate equality holds
(see Section~1.2 of the supplementary material [\citet{suppl}]) for the sample size in
question, that is, $n_2$:
%
\begin{equation}
\label{eq:informationloss} n_2  = \biggl( \biggl(\frac{c_{1}}{c_{2}}
\biggr)^{1+\alpha} { \biggl(\frac
{\ln c_{1}}{\ln c_{2}} \biggr)}^{2} {
\biggl(\frac{ c_2 - {c_2}^{\alpha}
}{ c_1 - {c_1}^{\alpha} } \biggr)}^{2} \biggr) n_{1}.
\end{equation}

\begin{figure}
\includegraphics{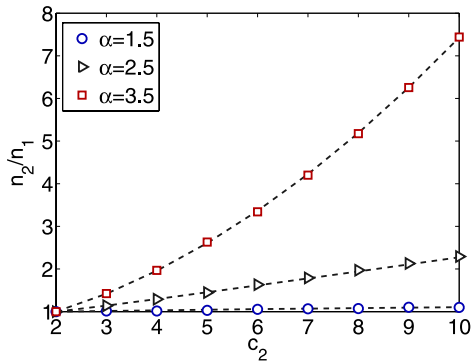}
\caption{The size of a data set required to achieve the same
statistical certainty in $\alpha$ (constant MSE) when using a coarser
binning scheme $c_{2}$, for several choices of $\alpha$. Since our
claim here is asymptotic, fixing a large value for $n_1$ (e.g., $10^4$)
results in a good agreement between the data points that come from a
simulation study and the dashed lines obtained analytically from
equation \protect\eqref{eq:informationloss}}
\label{fig:n1n2}
\end{figure}

Figure~\ref{fig:n1n2} illustrates how $n_2$ varies with the coarseness
of the second binning scheme $c_{2}$. For concreteness, we fix
$c_{1}=2$ and show the constant's behavior for several choices of
$\alpha$ and for schemes $c_{2}\geq2$. As expected, increasing the
coarseness of the binning scheme decreases the information available
for estimation, and the required sample size increases. Information
loss also arises from variation in $\alpha$. As $\alpha$ increases, the
variance of the generating distribution decreases, and a given sample
size will span fewer bins. The fundamental source of information loss
for estimation is the loss of bins, that is, the commingling of
observations that are distinct, which may arise either from coarsening
the binning scheme or from decreasing the variance of the generating
distribution.

%
%

%
\begin{figure}[b]
\includegraphics{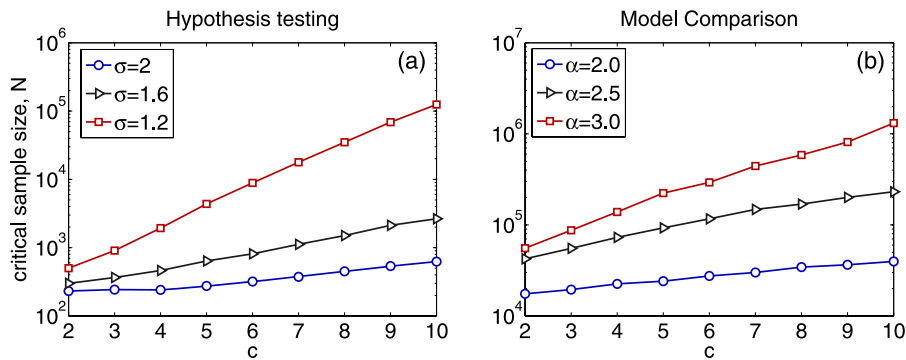}
\caption{Impact of information loss on: \textup{(a)} mean $p$-value using
synthetic log-normal data and \textup{(b)}~normalized log-likelihood ratio
$n^{-1/2} \mathcal{R}/\sigma$ using synthetic power-law data. All data
sets are logarithmically binned in powers of $c$ shown on the $x$-axis.
We show the critical value of the sample size, $N$, required to make
the correct decision for both these steps of our framework on the $y$-axis.}
\label{fig:infolossLRPV}
\end{figure}
%

The information-loss effect is sufficiently strong that a powers-of-10
binning scheme can require nearly eight times as large a sample to
obtain the same statistical accuracy in $\alpha$, when $\alpha>3$.
Thus, if the option is available during the experimental design phase
of a study, as fine a grained binning scheme as is possible should be
used in collecting the data in order to maximize subsequent statistical
accuracy.



We now illustrate the impact of loss of bins on the hypothesis testing
and the model comparison steps of our framework, respectively, using
two experiments: one in which we draw a sample from a log-normal
distribution with $\mu=0.3$ and $\sigma= \{1.2, 1.6, 2\}$ and second
in which we draw a sample from the power-law distribution with $\alpha
=\{2, 2.5, 3\}$. We keep $\bmin$ fixed for both the experiments. In the
first experiment, we test the plausibility of the power-law hypothesis
by computing the mean $p$-value and in the second experiment, we
compare the power-law and the log-normal hypotheses by computing the
normalized log-likelihood ratio.

We show the critical sample size, $N$, required to reject the power-law
hypothesis in the first experiment [Figure~\ref{fig:infolossLRPV}(a)] and to
favor the power-law hypothesis in the second experiment
[Figure~\ref{fig:infolossLRPV}(b)] as a function of the binning scheme $c$. We also
study the effect of variance of the underlying distribution by showing
the trend lines for $\sigma$ and $\alpha$, respectively, for the two
figures. Here again we observe that decreasing the variance (i.e.,
decreasing $\sigma$ for the log normal or increasing $\alpha$ for the
power law) results in loss of bins and, hence, larger $N$ is required
to reliably make the correct decision. Increasing $c$ has a similar
effect. Note that decreasing variance corresponds to the rate at which
$N$ has to be increased for making correct decisions.


\section{Applications to real-world data}
\label{sec:applications}
Having described statistically principled methods for working with
power-law distributions and binned empirical data, we now apply them to
analyze several real-world binned data sets to determine which of them
do and do not follow power-law distributions. As we will see, the
results indicate that some of these quantities are indeed consistent
with the power-law hypothesis, while others are not.

The 12 data sets we study are drawn from a broad variety of scientific
domains, including medicine, genetics, geology, ecology, meteorology,
earth sciences, demographics and the social sciences. They are as follows:
\begin{longlist}[10.]
\item[1.]
Estimated number of personnel in a terrorist
organization \citetp{asal:rethemeyer:2008}, binned by powers of ten,
expect that the first two bins are merged.

\item[2.] Diameter of branches in the plant species \textit
{Cryptomeria} \citetp{shinokazi:etal:1964}, binned in 30~mm intervals.

\item[3.] Volume of ice in an iceberg calving event \citetp
{chapuis:tetzlaff:2011}, binned by powers of ten.

\item[4.] Length of a patient's hospital stay within a year \citetp
{healthheritage11}, arbitrarily binned as natural numbers from 1 to 15,
plus one bin spanning 16--365 days. (Stays of length 0 are omitted.)

\item[5.] Wind speed (mph) of a tornado in the United States from 2007 to
2011 \citetp{spc11}, binned into categories according to the Enhanced
Fujita (EF) scale, a roughly logarithmic binning scheme.\footnote{Tornado data spanning 1950--2006, binned using the deprecated
Fujita scale, are also available. Repeating our analysis on these
yields the same conclusions.}

\item[6.]
Maximum wind speed (knots) of tropical
storms and hurricanes in the United States between 1949 and 2010
\citetp
{nhc11}, binned in 5-knot intervals.

\item[7.]
The human population of U.S. cities in the
2000 U.S. Census.

\item[8.]
Size (acres) of wildfires occurring on
U.S. federal land from 1986--1996 \citetp{newman:2005}.

\item[9.]
Intensity of earthquakes occurring in
California from 1910--1992, measured as the maximum amplitude of motion
during the quake \citetp{newman:2005}.

\item[10.]
Area (sq. km) of glaciers in
Scandinavia \citetp{wgi12}.

\item[11.] Number of cases per 100,000 of various rare diseases
\citetp
{prevalence11}.

\item[12.]
Number of genes associated with a
disease \citetp{goh07}.
\end{longlist}

%
\begin{sidewaystable}
\tablewidth=\textwidth
\caption{Details of the data sets described in Section \protect\ref
{sec:applications}, along with their power-law fits and the
corresponding $p$-values (\textbf{bold} values indicate statistically
plausible fits). $N$ denotes the full sample size, while $n$ is the
size of the fitted power-law region. Cases where we additionally
considered a restricted power-law fit (see text), with fixed $\bmin
=b_{1}$, are denoted by $\circ$ next to the $\bminhat$ value. Standard
error (std. err.) estimates were derived from a bootstrap using 1000
replications}
\label{tab:resplfit}
\begin{tabular*}{\textwidth}{@{\extracolsep{\fill}}lccd{2.3}ccccc@{}}
\hline
\textbf{Quantity} & $\bolds{N}$ & \textbf{Binning scheme} $\bolds{B}$ &
\multicolumn{1}{c}{$\bolds{\hat{\alpha}}$ }
& \textbf{Std. err.} & &
$\bolds{\bminhat}$ & $\bolds{n}$\textbf{, tail }& $\bolds{p (\pm0.03)}$ \\
\hline
Personnel in a terrorist group & 393 & logarithmic, $c=10$ &1.75 &
(0.11)\phantom{0} & & 1000 & 56 & \textbf{0.13} \\
--- & & & 1.29 & (0.01)\phantom{0} & $\circ$ & 1 & 393 & 0.00 \\
Plant branch diameter (mm) & 3897 & linear, 30 mm & 2.34 & (0.02)\phantom{0} & &
0.3 & 3897 & 0.00\\
Volume in iceberg calving ($\times10^{3}$~m${}^{3}$) & 5837 &
arbitrary & 1.29 & (0.02)\phantom{0} & & $1.26\times10^{12}$ & 143 & \textbf{0.49}
\\
--- & & & 1.155 & (0.002) & $\circ$ & $10.97$ & 5837 & 0.00 \\
Length of hospital stay & 11,769 & arbitrary\tabnoteref{tt21} & 3.24 & (0.27)\phantom{0} & & 14 & 303 & \textbf{0.40} \\
--- & & & 2.020 & (0.007) & $\circ$ & 1 & 11,769 & 0.00 \\
Wind speed, tornado (mph) & 7231 & EF-scale\tabnoteref{tt22}
& 7.10 & (0.20)\phantom{0} & & 111 & 980 & 0.03 \\
--- & & & 4.58 & (0.03)\phantom{0} & $\circ$ & 65 & 7231 & 0.00 \\
Max. wind speed, hurricane (knots) & 879 & linear, 5 knots & 14.20 &
(1.69)\phantom{0} & & 122.5 & 56 & \textbf{0.36} \\
--- & & & 2.44 & (0.03)\phantom{0} & $\circ$ & 32.5 & 879 & 0.00 \\
Population of city & 19,447 & logarithmic, $c=2$ & 2.38 & (0.07)\phantom{0} & &
65,536 & 426 & \textbf{0.72} \\
Size of wildfire (acres) & 203,785 & logarithmic, $c=2$ & 1.482 &
(0.002) & & 2 & 52,004 & 0.00 \\
Intensity of earthquake & 19,302 & logarithmic, $c=10$ & 1.82 & (0.02)\phantom{0}
& & 10,000 & 2659 & \textbf{0.18} \\
Size of glacier (km${}^2$) & 2428 & logarithmic, $c=2$ & 1.95 & (0.04)\phantom{0}
& & 1 & 635 & 0.04 \\
Rare disease prevalence & 675 & logarithmic, $c=2$ & 2.88 & (0.14)\phantom{0} & &
16 & 99 & 0.00 \\
Genes associated with disease & 1284 & logarithmic, $c=2$ & 2.72 &
(0.12)\phantom{0} & & 8 & 217 & \textbf{0.87} \\
--- & & & 1.75 & (0.01)\phantom{0} & $\circ$ & 1 & 1284 & 0.00\\
\hline
\end{tabular*}
\tabnotetext[a]{tt21}{\citetp{healthheritage11}.}
\tabnotetext[b]{tt22}{\citetp{spc11}.}
\end{sidewaystable}
%

Data sets 1--6 are naturally binned,
that is, bins are fixed as given and either the raw observations are
unavailable or analyses of such data typically focus on binned
observations. Raw values for data sets 7--12 are available, and these quantities are included for other
reasons. Data sets 7--9 were also
analyzed in \citet{clauset:shalizi:newman:2009}, and we reanalyze them
in order to illustrate that similar conclusions may be extracted
despite binning or to highlight differences induced by binning. Data
sets 10--12 were analyzed as binned data
by their primary sources, and we do the same to ensure comparability of
our results.

\begin{figure}
\includegraphics{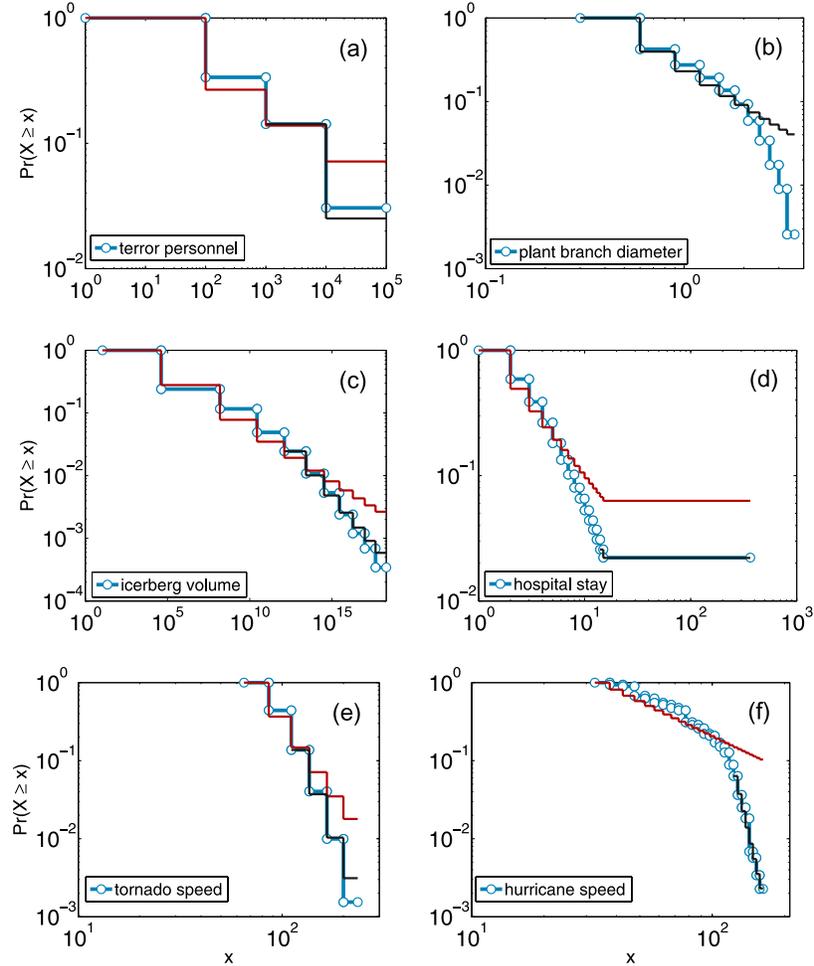}
\caption{Empirical distributions (as complementary c.d.f.s) $\Pr(X\geq x)$
for data sets 1--6: the \textup{(a)}~number of
personnel in a terrorist organization, \textup{(b)} diameter of branches in
plants of the species \textit{Cryptomeria}, \textup{(c)} volume of ice in an
iceberg calving event, \textup{(d)} length of a patient's hospital stay,
\textup{(e)}~wind speed of tornados, and \textup{(f)} maximum wind speed of tropical storms
and hurricanes, along with the best fitting power-law distribution with
$\bmin$ estimated (black) and $\bmin$ fixed at the smallest bin
boundary (red).}
\label{fig:empfits1}
\end{figure}

\begin{figure}[t]
\includegraphics{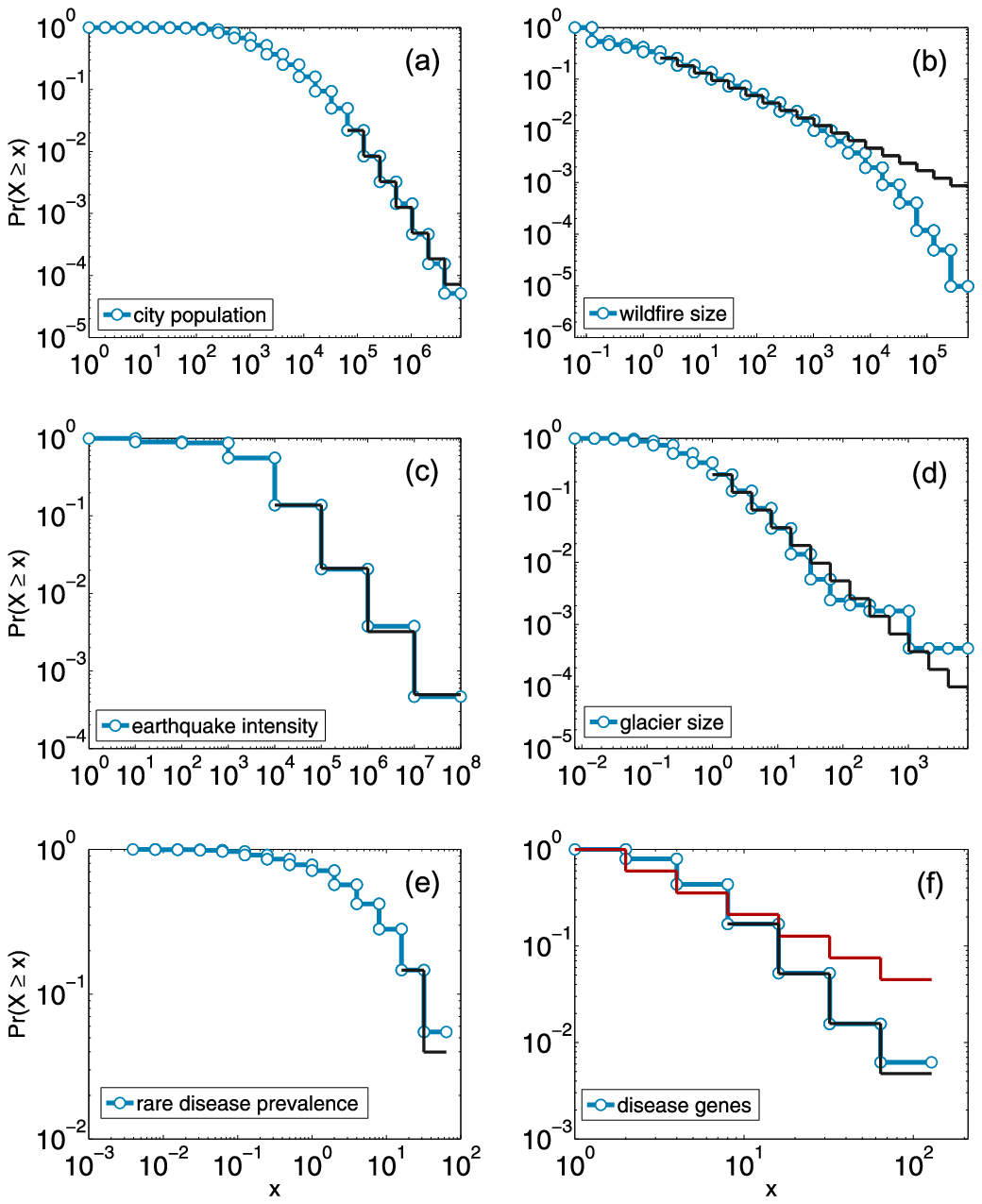}
\caption{Empirical distributions (as complementary c.d.f.s) $\Pr(X\geq x)$
for data sets 7--12:
the \textup{(a)}~human
population of U.S. cities, \textup{(b)} size of wildfires on U.S. federal
land, \textup{(c)} intensity of earthquakes in California, \textup{(d)} area of glaciers
in Scandanavia, \textup{(e)} prevalence of rare diseases, and \textup{(f)} number of
genes associated with a disease, along with the best fitting power-law
distribution with $\bmin$ estimated (black) and $\bmin$ fixed at the
smallest bin boundary (red).}
\label{fig:empfits2}
\end{figure}

Table~\ref{tab:resplfit} summarizes each data set and gives the
parameters of the best fitting power law. Figures~\ref{fig:empfits1}
and \ref{fig:empfits2} plot the empirical bin counts and the fitted
power-law models. In several cases, we also include fits where we have
fixed $\bmin=b_{1}$, the smallest bin boundary in order to test the
power-law model on the entire data set. This supplementary test was
conducted when either a previous claim had been made regarding the
entire distribution's shape or when visual inspection suggested that
such a claim might be reasonable. Finally, Table~\ref{tab:rescomp}
summarizes the results of the likelihood ratio tests and includes our
judgement of the statistical support for the power-law hypothesis with
each data set.

For none of the quantities was the power-law hypothesis strongly
supported, which requires that the power law was both a good fit to the
data and a better fit than the alternatives. This fact reinforces the
difficulty of distinguishing genuine power-law behavior from
nonpower-law-but-still-heavy-tailed behavior. In most cases, the
likelihood ratio test against the exponential distribution confirms the
heavy-tailed nature of these quantities, that is, the power law was
typically a better fit than the exponential, except for the length of
hospital stays, tornado wind speeds and the prevalences of rare diseases.

Two quantities---the number of personnel in a terrorist organization
and the number of genes associated with a disease---yielded weak
support for the power-law hypothesis, in which the power law was a good
fit, but at least one alternative was better. In the case of the
gene-disease data, this quantity is better fit by a log-normal
distribution, suggesting some kind of multiplicative stochastic process
as the underlying mechanism. The terror personnel data is better fit by
both the log-normal and the stretched exponential distributions,
however, given that so few observations ended up in the tail region,
the case for any particular distribution is not strong.

%
%

%

%

Five quantities produced moderate support for the power law hypothesis,
in which the power law was a good fit but alternatives like the
log-normal or stretched exponential remain plausible, that is, their
likelihood ratio tests were inconclusive. In particular, the volume of
icebergs, the length of hospital stays (but see above), the maximum
wind speed of a hurricane, the population of a city and the intensity
of earthquakes all have moderate support.

Of the six supplemental tests we conducted, in which we fixed $\bmin
=b_{1}$, only two---the maximum wind speed of hurricanes and the size
of wildfires---yielded any support for a power law, and in both cases
the power-law distribution with exponential cutoff was better than the
pure power law. In the case of hurricanes, a~cutoff is scientifically
reasonable: windspeed in hurricanes is related to their spatial size,
which is ultimately constrained by the size of convection cells in the
upper atmosphere, the distribution of the continents and the rate at
which energy is transferred from the ocean surface \citetp
{persing:montgomery:2003}. The presence of such physical constraints
implies that any scale invariance present in the underlying generative
process must be truncated by the finite size of the Earth itself.
Larger planets, like Jupiter and Saturn, may thus exhibit scaling over
a larger range of storm intensities, although we know of no systematic
data set of extraterrestrial storms.

%
\begin{sidewaystable}
\tablewidth=\textwidth
\caption{Comparison of the fitted power-law behavior against
alternatives. For each data set, we give the power law's $p$-value from
Table \protect\ref{tab:resplfit}, the log-likelihood ratios against
alternatives, and the $p$-value for the significance of each likelihood
ratio test. Statistically significant values are given in
\textbf{bold}. Positive log-likelihood ratios indicate that the power law is
favored over the alternative. For nonnested alternatives, we report
the normalized log-likelihood ratio $n^{-1/2}\mathcal{R}\sigma$; for
nested models (the power law with exponential cutoff), we give the
actual log-likelihood ratios. The final column lists our judgement of
the statistical support for the power-law hypothesis with each data
set. ``None'' indicates data sets that are probably not power-law
distributed; ``weak'' indicates that the power law is a good fit but a
nonpower law alternative is better; ``moderate'' indicates that the
power law is a good fit but alternatives remain plausible. No quantity
achieved a ``good'' label, where the power law is a good fit and none
of the alternatives is considered plausible. In some cases, we write
``with cutoff'' to indicate that the power law with exponential cutoff
is clearly favored over the pure power law. In each of these cases,
however, some of the alternatives are also good fits, such as the
log-normal and stretched exponential}
\label{tab:rescomp}
\begin{tabular*}{\textwidth}{@{\extracolsep{\fill}}lcd{2.2}cd{2.2}cd{2.2}cd{3.2}cc@{}}
\hline
& & \multicolumn{2}{c} {\textbf{Log-normal}} & \multicolumn{2}{c}
{\textbf{Exponential}} & \multicolumn{2}{c}{\textbf{Stretched exp.}} & \multicolumn
{2}{c@{}}{\textbf{Power law}${} \bolds{+} {}$\textbf{cutoff}} & \\[-6pt]
&  & \multicolumn{2}{c} {\hrulefill} & \multicolumn{2}{c}
{\hrulefill} & \multicolumn{2}{c}{\hrulefill} & \multicolumn
{2}{c@{}}{\hrulefill} & \\
\multicolumn{1}{@{}l}{\textbf{Quantity}} &
\multicolumn{1}{c}{\textbf{Power law} $\bolds{p}$} & \multicolumn{1}{c}{\textbf{LR}} &
\multicolumn{1}{c}{$\bolds{p}$} & \multicolumn{1}{c}{\textbf{LR}} &
\multicolumn{1}{c}{$\bolds{p}$} & \multicolumn{1}{c}{\textbf{LR}} &
\multicolumn{1}{c}{$\bolds{p}$} & \multicolumn{1}{c}{\textbf{LR}} &
\multicolumn{1}{c}{$\bolds{p}$} &
\multicolumn{1}{c@{}}{\textbf{Support for power law}}\\
\hline
Personnel in a terrorist group & \textbf{0.13} & \mbox{ }-2.01 &
\textbf{{0.04}} & 3.91 & \textbf{{0.00}} & -1.93 & \textbf{{0.05}} &
-2.57 & 0.11 & weak\\
--- & 0.00 & -4.32 & \textbf{{0.00}} & 4.59 & \textbf{{0.00}}
& -4.47 & \textbf{{0.00}} & -26.26 & \textbf{{0.00}} & none\\
Plant branch diameter & 0.00 & -9.71 & \textbf{{0.00}} &
1.99 & \textbf{{0.05}} & -9.48 & \textbf{{0.00}} & -123.8 & \textbf{{0.00}}& none \\
Volume in iceberg calving & \textbf{0.49} & -1.12 & 0.26 &
10.61 & \textbf{{0.00}} & -1.16 & 0.24 & -1.70 &
0.19 & moderate\\
--- & 0.00 & 0.85 & 0.40 & 43.02 & \textbf{{0.00}} & \mbox{ } 2.26 & \textbf{{0.01}} & -13.29 & \textbf{{0.00}} & none \\
Length of hospital stay & \textbf{0.40} &
-0.98&0.33&-1.02&0.31&-1.01&0.31&-0.23&0.63& moderate \\
--- & 0.00 & -18.37&\textbf{{0.00}}&-1.86&\textbf{{0.06}}&-18.69&\textbf{{0.00}}&-602.9&\textbf{{0.00}}& none \\
Wind speed, tornado & 0.03 & -3.16 & \textbf{{0.00}} & -3.32 & \textbf{{0.00}} & -2.72 & \textbf{{0.01}} & -7.92 &
\textbf{{0.01}} & none \\
--- & 0.00 & -17.36 & \textbf{{0.00}} & -19.22 & \textbf{{0.00}} &
-13.85 &
\textbf{{0.00}} & -214.6 & \textbf{{0.00}} & none\\
Max. wind speed, hurricane & \textbf{0.36} &-0.35&0.73& 6.17&\textbf{{0.00}}&-0.72&0.48&-0.30&0.6\phantom{0}& moderate \\
--- &0.00&-13.26&\textbf{{0.00}}&-20.71&\textbf{{0.00}}&-13.78&\textbf{{0.00}}&-117.1&\textbf{{0.00}}& with cutoff \\
Population of city & \textbf{0.72} &-0.07 & 0.95 & 16.25 & \textbf{{0.00}} & -0.08 & 0.94 & -0.23 & 0.63 &
moderate\\
Size of wildfire & 0.00 & -16.03 & \textbf{{0.00}} & 9.26& \textbf{{0.00}} & -16.42 & \textbf{{0.00}}& -410 & \textbf{{0.00}} & with
cutoff\\
Intensity of earthquake & \textbf{0.18} & 1.02 & 0.27 & \mbox{ }21.63 & \textbf{0.00} & 0.75 & 0.45 & -0.78 & 0.38
& moderate \\
Size of glacier & 0.04 &-0.56 & 0.58 & 1.01 & 0.31 &
-0.56 & 0.58 & 0.00 & 0.96 & none \\
Rare disease prevalence & 0.00 & -4.72 & \textbf{{0.00}} & -4.64 &
\textbf{{0.00}} & -3.77 & \textbf{{0.00}} & -7.55 &
\textbf{{0.01}} & none\\
Genes associated with disease & \textbf{0.87} & -2.52 & \textbf{{0.01}} & 2.92 & \textbf{{0.00}} & -0.49 & 0.63 & -0.51 & 0.48 & weak
\\
--- & 0.00 & -11.28 & \textbf{{0.00}} & -3.14 & \textbf{{0.00}} &
-10.83 & \textbf{{0.00}} & -159.4 & \textbf{{0.00}} & none
\\
\hline
\end{tabular*}
\end{sidewaystable}
%

For the three data sets also analyzed in \citet
{clauset:shalizi:newman:2009}---city populations, wildfire sizes and
earthquake intensities---we reassuringly come to similar conclusions
when analyzing their binned counterparts. The one exception is the
intensity of earthquakes, which illustrates the impact of information
loss from binning. The first consequence is that our choice $\bmin$ is
slightly larger than the $\xmin$ estimated from the raw data. The
slight curvature in this distribution's tail means this difference
raises our scaling parameter estimate to $\hat{\alpha}=1.82\pm0.02$
compared to $\hat{\alpha}=1.64\pm0.04$ in \citet
{clauset:shalizi:newman:2009}.
Furthermore, \citet{clauset:shalizi:newman:2009} found the power law to
be a poor fit by itself ($p=0.00\pm0.03$) and that the power-law with a
cutoff was heavily favored. In contrast, we failed to reject the power
law ($p=0.18\pm0.03$) and the comparison to the power law with cutoff
was inconclusive. That is, the information lost by binning obscured the
more clearcut results obtained on raw data for earthquake intensities.

In some cases, our conclusions have direct implications for theoretical
work, shedding immediate light on what type of theoretical explanations
should or should not be considered for the corresponding phenomena. An
illustrative example is the branch diameter data. Past work on the
branching structure of plants \citetp
{yamamoto93,shinokazi:etal:1964,west09} has argued for a fractal model,
in which certain conservation laws imply a power-law distribution for
branch diameters within a plant. Some theories go further, arguing that
a forest is a kind of a ``scaled up'' plant and that the power-law
distribution of branch diameters extends to entire collections of
naturally co-occuring plants. Critically, the branch data analyzed
here, and its purported power-law shape, have been cited as evidence
supporting these claims \citetp{west09}. However, our results show that
these data provide no statistical support for the power-law hypothesis
[we find similar results for the other binned data of \citet
{shinokazi:etal:1964,west09}]. Our results thus demonstrate that these
theories' predictions do not match the empirical data and alternative
explanations should be considered. Indeed, our results suggest that the
basic pipe model itself is flawed or incomplete, as we observe too few
large-diameter branches and too many small-diameter branches compared
to the theory's prediction.

In other cases, our results suggest specific theoretical processes to
be considered. For instance, the full distribution of hospital stays is
better fit by all the alternative distributions than by the power law,
but the stretched exponential is of particular interest. Survival
analysis is often framed in terms of hazard rates, that is, a Poisson
process with a nonstationary event probability, and our results suggest
that such a model may be worth considering: if the hazard rate for
leaving the hospital decreases as the length of the stay increases, a
heavy-tailed distribution like the stretched exponential is produced.
Additional investigation of the covariates that best predict the
trajectory of this hazard rate would provide a test of this hypothesis.

\section{Conclusions}
\label{sec:conclusions}

The primary goal of this article was to introduce a principled
framework for testing the power-law hypothesis with binned empirical
data, based on the framework introduced in \citet
{clauset:shalizi:newman:2009}, and to explore the impact of information
loss due to binning on the resulting statistical conclusions. Although
the information loss can be severe for coarse binning schemes, for
example, powers-of-10, sound statistical conclusions can still be made
using these methods. These methods should allow practitioners in a wide
variety of fields to better distinguish power-law from nonpower-law
behavior in empirical data, regardless of whether the data are binned
or not.


In applying our methods to a large number of data sets from various
fields, we found that the data for many of these quantities are not
compatible with the hypothesis that they were drawn from a power-law
distribution. In a few cases, the data were found to be compatible, but
not fully: in these cases, there was ample evidence that alternative
heavy-tailed distributions are an equally good or better explanation.

The study of power laws is an exciting effort that spans many
disciplines, and their identification in complex systems is often
interpreted as evidence for, or suggestions of, theoretically
interesting processes. In this paper, we have argued that the common
practice of identifying and quantifying power-law distributions by the
approximately straight-line behavior on a binned histogram on a doubly
logarithmic plot should not be trusted: such straight-line behavior is
a necessary but not sufficient condition for true power-law behavior.
Furthermore, binned data present special problems because conventional
methods for testing the power-law hypothesis \citetp
{clauset:shalizi:newman:2009} could only be applied to continuous or
integer-valued observations. By extending these techniques to binned
data, we enable researchers to reliably investigate the power-law
hypothesis even when the data do not take a convenient form, either
because of the way they were collected, because the original values are
lost, or for some other reason.

Properly applied, these methods can provide objective evidence for or
against the claim that a particular distribution follows a power law.
(In principle, our binned methods could be extended to other,
nonpower-law distributions, although we do not provide such extensions
here.) Such objective evidence provides statistical rigor to the larger
goal of identifying and characterizing the underlying processes that
generate these observed patterns. That being said, answers to some
questions of scientific interest may not depend solely on the
distribution following a power law perfectly. Whether or not a quantity
not following a power law poses a problem for a researcher depends
largely on his or her scientific goals, and in some cases a power law
may not be more fundamentally interesting than some other heavy-tailed
distribution such as the log-normal or the stretched-exponential.

In closing, we emphasize that the identification of a power law in some
data is only part of the challenge we face in explaining their causes
and implications in natural and man-made phenomena. We also need
methods by which to test the processes proposed to explain the observed
power laws and to leverage these interesting patterns for practical
purposes. This perspective has a long and ongoing history, reaching at
least as far back as \citet{ijiri:simon:1977}, with modern analogs
given by \citet{mitzenmacher06} and by \citet{stumpf:porter:2012}. We
hope the statistical tools presented here aid in these endeavors.

\section*{Acknowledgments}
The authors thank Cosma Shalizi and Daniel Larremore for helpful
conversations, Amy Wesolowski for contributions to an earlier version
of this project, and Luke Winslow, Valentina Radic, Anne Chapius, Brian
Enquist and Victor Asal for sharing data.
Implementations of our numerical methods are available online at
\texttt{\href{http://www.santafe.edu/\textasciitilde aaronc/powerlaws/bins/}{http://www.santafe.edu/}}
\texttt{\href{http://www.santafe.edu/\textasciitilde aaronc/powerlaws/bins/}{\textasciitilde aaronc/powerlaws/bins/}}.

%


\begin{supplement}[id=suppA]
\stitle{Supplement to ``Power-law distributions in binned empirical data''}
\slink[doi]{10.1214/13-AOAS710SUPP} 
\sdatatype{.pdf}
\sfilename{aoas710\_supp.pdf}
\sdescription{In this supplemental file, we derive a closed-form
expression for the binned MLE in Section~1.1, quantify the amount of
information loss on using a coarser binning scheme in Section~1.2 and
include the likelihood ratio test for the binned case in Section~2.}
\end{supplement}

%
%

%

\printaddresses


\begin{thebibliography}{55}


\bibitem[\protect\citeauthoryear{Aban and Meerschaert}{2004}]{aban:meerschaert:2004}
\begin{barticle}[mr]
\bauthor{\bsnm{Aban},~\bfnm{Inmaculada~B.}\binits{I.~B.}} \AND
\bauthor{\bsnm{Meerschaert},~\bfnm{Mark~M.}\binits{M.~M.}}
(\byear{2004}).
\btitle{Generalized least-squares estimators for the thickness of heavy tails}.
\bjournal{J. Statist. Plann. Inference}
\bvolume{119}
\bpages{341--352}.
\bid{doi={10.1016/S0378-3758(02)00419-6}, issn={0378-3758}, mr={2019645}}
\end{barticle}
\bptok{imsref}%
\endbibitem

\bibitem[\protect\citeauthoryear{Arnold}{1983}]{arnold:1983}
\begin{bbook}[mr]
\bauthor{\bsnm{Arnold},~\bfnm{Barry~C.}\binits{B.~C.}}
(\byear{1983}).
\btitle{Pareto Distributions}.
\bseries{Statistical Distributions in Scientific Work}
\bvolume{5}.
\bpublisher{International Co-operative Publishing House},
\blocation{Burtonsville, MD}.
\bid{mr={0751409}}
\end{bbook}
\bptok{imsref}%
\endbibitem

\bibitem[\protect\citeauthoryear{Asal and Rethemeyer}{2008}]{asal:rethemeyer:2008}
\begin{barticle}[author]
\bauthor{\bsnm{Asal},~\bfnm{V.}\binits{V.}} \AND
\bauthor{\bsnm{Rethemeyer},~\bfnm{R.~K.}\binits{R.~K.}}
(\byear{2008}).
\btitle{The nature of the beast: Organizational structures and the lethality of terrorist attacks}.
\bjournal{The Journal of Politics}
\bvolume{70}
\bpages{437--449}.
\end{barticle}
\bptok{imsref}%
\endbibitem

\bibitem[\protect\citeauthoryear{Barndorff-Nielsen and Cox}{1995}]{barndorffnielsen:cox:1995}
\begin{bbook}[author]
\bauthor{\bsnm{Barndorff-Nielsen},~\bfnm{O.~E.}\binits{O.~E.}} \AND
\bauthor{\bsnm{Cox},~\bfnm{D.~R.}\binits{D.~R.}}
(\byear{1995}).
\btitle{Inference and Asymptotics}.
\bpublisher{Chapman \& Hall},
\blocation{London}.
\end{bbook}
\bptok{imsref}%
\endbibitem

\bibitem[\protect\citeauthoryear{Beirlant and Teugels}{1989}]{beirlant:teugels:1989}
\begin{barticle}[author]
\bauthor{\bsnm{Beirlant},~\bfnm{J.}\binits{J.}} \AND
\bauthor{\bsnm{Teugels.},~\bfnm{J.~L.}\binits{J.~L.}}
(\byear{1989}).
\btitle{Asymptotic normality of Hill's estimator}.
\bjournal{Extreme Value Theory}
\bvolume{51}
\bpages{148--155}.
\end{barticle}
\bptok{imsref}%
\endbibitem

\bibitem[\protect\citeauthoryear{Breiman, Stone and Kooperberg}{1990}]{breiman:stone:kooperberg:1990}
\begin{barticle}[mr]
\bauthor{\bsnm{Breiman},~\bfnm{Leo}\binits{L.}},
\bauthor{\bsnm{Stone},~\bfnm{Charles~J.}\binits{C.~J.}} \AND
\bauthor{\bsnm{Kooperberg},~\bfnm{Charles}\binits{C.}}
(\byear{1990}).
\btitle{Robust confidence bounds for extreme upper quantiles}.
\bjournal{J. Stat. Comput. Simul.}
\bvolume{37}
\bpages{127--149}.
\bid{doi={10.1080/00949659008811300}, issn={0094-9655}, mr={1082452}}
\end{barticle}
\bptok{imsref}%
\endbibitem

\bibitem[\protect\citeauthoryear{Cadez et~al.}{2002}]{cadez:etal:2002}
\begin{barticle}[author]
\bauthor{\bsnm{Cadez},~\bfnm{I.~V.}\binits{I.~V.}},
\bauthor{\bsnm{Smyth},~\bfnm{P.}\binits{P.}},
\bauthor{\bsnm{McLachlan},~\bfnm{G.~J.}\binits{G.~J.}} \AND
\bauthor{\bsnm{McLaren},~\bfnm{C.~E.}\binits{C.~E.}}
(\byear{2002}).
\btitle{Maximum likelihood estimation of mixture of densities for binned and truncated multivariate data}.
\bjournal{Machine Learning}
\bvolume{47}
\bpages{7--34}.
\end{barticle}
\bptok{imsref}%
\endbibitem

\bibitem[\protect\citeauthoryear{Chapuis and Tetzlaff}{2012}]{chapuis:tetzlaff:2011}
\begin{bmisc}[author]
\bauthor{\bsnm{Chapuis},~\bfnm{A.}\binits{A.}} \AND
\bauthor{\bsnm{Tetzlaff},~\bfnm{T.}\binits{T.}}
(\byear{2012}).
\bhowpublished{The variability of tidewater-glacier calving: Origin of event-size and interval distributions.
Available at \arxivurl{arXiv:1205.1640}.}
\end{bmisc}
\bptok{imsref}%
\endbibitem

\bibitem[\protect\citeauthoryear{Clauset, Shalizi and Newman}{2009}]{clauset:shalizi:newman:2009}
\begin{barticle}[mr]
\bauthor{\bsnm{Clauset},~\bfnm{Aaron}\binits{A.}},
\bauthor{\bsnm{Shalizi},~\bfnm{Cosma~Rohilla}\binits{C.~R.}} \AND
\bauthor{\bsnm{Newman},~\bfnm{M.~E.~J.}\binits{M.~E.~J.}}
(\byear{2009}).
\btitle{Power-law distributions in empirical data}.
\bjournal{SIAM Rev.}
\bvolume{51}
\bpages{661--703}.
\bid{doi={10.1137/070710111}, issn={0036-1445}, mr={2563829}}
\end{barticle}
\bptok{imsref}%
\endbibitem

\bibitem[\protect\citeauthoryear{Clauset and Woodard}{2013}]{clauset:woodard:2012}
\begin{barticle}[author]
\bauthor{\bsnm{Clauset},~\bfnm{A.}\binits{A.}} \AND
\bauthor{\bsnm{Woodard},~\bfnm{R.}\binits{R.}}
(\byear{2013}).
\btitle{Estimating the historical and future probabilities
of large terrorist events}.
\bjournal{Ann. Appl. Stat.}
\bvolume{7}
\bpages{1838--1865}.
\end{barticle}
\bptok{imsref}%
\endbibitem

\bibitem[\protect\citeauthoryear{Clauset, Young and Gleditsch}{2007}]{clauset:etal:2007}
\begin{barticle}[author]
\bauthor{\bsnm{Clauset},~\bfnm{A.}\binits{A.}},
\bauthor{\bsnm{Young},~\bfnm{M.}\binits{M.}} \AND
\bauthor{\bsnm{Gleditsch},~\bfnm{K.~S.}\binits{K.~S.}}
(\byear{2007}).
\btitle{On the frequency of severe terrorist events}.
\bjournal{Journal of Conflict Resolution}
\bvolume{51}
\bpages{58--87}.
\end{barticle}
\bptok{imsref}%
\endbibitem

\bibitem[\protect\citeauthoryear{Cram{\'e}r}{1946}]{cramer46}
\begin{barticle}[mr]
\bauthor{\bsnm{Cram{\'e}r},~\bfnm{Harald}\binits{H.}}
(\byear{1946}).
\btitle{A contribution to the theory of statistical estimation}.
\bjournal{Skand. Aktuarietidskr.}
\bvolume{29}
\bpages{85--94}.
\bid{mr={0017505}}
\end{barticle}
\bptok{imsref}%
\endbibitem

\bibitem[\protect\citeauthoryear{Danielsson et~al.}{2001}]{danielsson:etal:2001}
\begin{barticle}[mr]
\bauthor{\bsnm{Danielsson},~\bfnm{J.}\binits{J.}},
\bauthor{\bparticle{de} \bsnm{Haan},~\bfnm{L.}\binits{L.}},
\bauthor{\bsnm{Peng},~\bfnm{L.}\binits{L.}} \AND
\bauthor{\bparticle{de} \bsnm{Vries},~\bfnm{C.~G.}\binits{C.~G.}}
(\byear{2001}).
\btitle{Using a bootstrap method to choose the sample fraction in tail index estimation}.
\bjournal{J. Multivariate Anal.}
\bvolume{76}
\bpages{226--248}.
\bid{doi={10.1006/jmva.2000.1903}, issn={0047-259X}, mr={1821820}}
\end{barticle}
\bptok{imsref}%
\endbibitem

\bibitem[\protect\citeauthoryear{Dekkers and de~Haan}{1993}]{dekkers:dehann:1993}
\begin{barticle}[mr]
\bauthor{\bsnm{Dekkers},~\bfnm{Arnold~L.~M.}\binits{A.~L.~M.}} \AND
\bauthor{\bparticle{de} \bsnm{Haan},~\bfnm{Laurens}\binits{L.}}
(\byear{1993}).
\btitle{Optimal choice of sample fraction in extreme-value estimation}.
\bjournal{J. Multivariate Anal.}
\bvolume{47}
\bpages{173--195}.
\bid{doi={10.1006/jmva.1993.1078}, issn={0047-259X}, mr={1247373}}
\end{barticle}
\bptok{imsref}%
\endbibitem

\bibitem[\protect\citeauthoryear{Drees and Kaufmann}{1998}]{drees:kaufmann:1998}
\begin{barticle}[mr]
\bauthor{\bsnm{Drees},~\bfnm{Holger}\binits{H.}} \AND
\bauthor{\bsnm{Kaufmann},~\bfnm{Edgar}\binits{E.}}
(\byear{1998}).
\btitle{Selecting the optimal sample fraction in univariate extreme value estimation}.
\bjournal{Stochastic Process. Appl.}
\bvolume{75}
\bpages{149--172}.
\bid{doi={10.1016/S0304-4149(98)00017-9}, issn={0304-4149}, mr={1632189}}
\end{barticle}
\bptok{imsref}%
\endbibitem

\bibitem[\protect\citeauthoryear{Efron and Tibshirani}{1993}]{efron:tibshirani:1993}
\begin{bbook}[mr]
\bauthor{\bsnm{Efron},~\bfnm{Bradley}\binits{B.}} \AND
\bauthor{\bsnm{Tibshirani},~\bfnm{Robert~J.}\binits{R.~J.}}
(\byear{1993}).
\btitle{An Introduction to the Bootstrap}.
\bseries{Monographs on Statistics and Applied Probability}
\bvolume{57}.
\bpublisher{Chapman \& Hall},
\blocation{New York}.
\bid{mr={1270903}}
\end{bbook}
\bptok{imsref}%
\endbibitem

\bibitem[\protect\citeauthoryear{Gabaix}{2009}]{gabaix:2009}
\begin{barticle}[author]
\bauthor{\bsnm{Gabaix},~\bfnm{X.}\binits{X.}}
(\byear{2009}).
\btitle{Power laws in economics and finance}.
\bjournal{Annual Review of Economics}
\bvolume{1}
\bpages{255--293}.
\end{barticle}
\bptok{imsref}%
\endbibitem

\bibitem[\protect\citeauthoryear{Goh et~al.}{2007}]{goh07}
\begin{barticle}[pbm]
\bauthor{\bsnm{Goh},~\bfnm{Kwang-Il}\binits{K.-I.}},
\bauthor{\bsnm{Cusick},~\bfnm{Michael~E.}\binits{M.~E.}},
\bauthor{\bsnm{Valle},~\bfnm{David}\binits{D.}},
\bauthor{\bsnm{Childs},~\bfnm{Barton}\binits{B.}},
\bauthor{\bsnm{Vidal},~\bfnm{Marc}\binits{M.}} \AND
\bauthor{\bsnm{Barab{\'{a}}si},~\bfnm{Albert-L{\'{a}}szl{\'{o}}}\binits{A.-L.}}
(\byear{2007}).
\btitle{The human disease network}.
\bjournal{Proc. Natl. Acad. Sci. USA}
\bvolume{104}
\bpages{8685--8690}.
\bid{doi={10.1073/pnas.0701361104}, issn={0027-8424}, pii={0701361104}, pmcid={1885563}, pmid={17502601}}
\end{barticle}
\bptok{imsref}%
\endbibitem

\bibitem[\protect\citeauthoryear{Goldstein, Morris and Yen}{2004}]{goldstein:morris:yen:2004}
\begin{barticle}[author]
\bauthor{\bsnm{Goldstein},~\bfnm{M.~L.}\binits{M.~L.}},
\bauthor{\bsnm{Morris},~\bfnm{S.~A.}\binits{S.~A.}} \AND
\bauthor{\bsnm{Yen},~\bfnm{G.~G.}\binits{G.~G.}}
(\byear{2004}).
\btitle{Problems with fitting to the power-law distribution}.
\bjournal{Eur. Phys. J. B}
\bvolume{41}
\bpages{255--258}.
\end{barticle}
\bptok{imsref}%
\endbibitem

\bibitem[\protect\citeauthoryear{Gr{\"{u}}nwald}{2007}]{grunwald:2007}
\begin{bbook}[author]
\bauthor{\bsnm{Gr{\"{u}}nwald},~\bfnm{P.~D.}\binits{P.~D.}}
(\byear{2007}).
\btitle{The Minimum Length Description Principle}.
\bpublisher{MIT Press},
\blocation{Cambridge, MA}.
\end{bbook}
\bptok{imsref}%
\endbibitem

\bibitem[\protect\citeauthoryear{Hall}{1982}]{hall:1982}
\begin{barticle}[mr]
\bauthor{\bsnm{Hall},~\bfnm{Peter}\binits{P.}}
(\byear{1982}).
\btitle{On some simple estimates of an exponent of regular variation}.
\bjournal{J. R. Stat. Soc. Ser.~B Stat. Methodol.}
\bvolume{44}
\bpages{37--42}.
\bid{issn={0035-9246}, mr={0655370}}
\end{barticle}
\bptok{imsref}%
\endbibitem

\bibitem[\protect\citeauthoryear{Handcock and Jones}{2004}]{handcock:jones:2004}
\begin{barticle}[author]
\bauthor{\bsnm{Handcock},~\bfnm{M.~S.}\binits{M.~S.}} \AND
\bauthor{\bsnm{Jones},~\bfnm{J.~H.}\binits{J.~H.}}
(\byear{2004}).
\btitle{Likelihood-based inference for stochastic models of sexual network evolution}.
\bjournal{Theoretical Population Biology}
\bvolume{65}
\bpages{413--422}.
\end{barticle}
\bptok{imsref}%
\endbibitem

\bibitem[\protect\citeauthoryear{{Heritage Provider Network}}{2012}]{healthheritage11}
\begin{bmisc}[author]
\borganization{Heritage Provider Network}
(\byear{2012}).
\bhowpublished{Health heritage prize data files, HHP\_release3.
Available at \url{http://bit.ly/wG8Psl}}.
\end{bmisc}
\bptok{imsref}%
\endbibitem

\bibitem[\protect\citeauthoryear{Hill}{1975}]{hill:1975}
\begin{barticle}[mr]
\bauthor{\bsnm{Hill},~\bfnm{Bruce~M.}\binits{B.~M.}}
(\byear{1975}).
\btitle{A simple general approach to inference about the tail of a distribution}.
\bjournal{Ann. Statist.}
\bvolume{3}
\bpages{1163--1174}.
\bid{issn={0090-5364}, mr={0378204}}
\end{barticle}
\bptok{imsref}%
\endbibitem

\bibitem[\protect\citeauthoryear{Horn}{1977}]{horn:1977}
\begin{barticle}[author]
\bauthor{\bsnm{Horn},~\bfnm{S.~D.}\binits{S.~D.}}
(\byear{1977}).
\btitle{Goodness-of-fit tests for discrete data: A review and an application to a health development scale}.
\bjournal{Biometrics}
\bvolume{33}
\bpages{237--247}.
\end{barticle}
\bptok{imsref}%
\endbibitem

\bibitem[\protect\citeauthoryear{Ijiri and Simon}{1977}]{ijiri:simon:1977}
\begin{bbook}[author]
\bauthor{\bsnm{Ijiri},~\bfnm{Y.}\binits{Y.}} \AND
\bauthor{\bsnm{Simon},~\bfnm{H.~A.}\binits{H.~A.}}
(\byear{1977}).
\btitle{Skew Distributions and the Sizes of Business Firms}.
\bpublisher{North-Holland},
\blocation{Amsterdam}.
\end{bbook}
\bptok{imsref}%
\endbibitem

\bibitem[\protect\citeauthoryear{Jarvinen, Neumann and Davis}{2012}]{nhc11}
\begin{bmisc}[author]
\bauthor{\bsnm{Jarvinen},~\bfnm{B.}\binits{B.}},
\bauthor{\bsnm{Neumann},~\bfnm{C.}\binits{C.}} \AND
\bauthor{\bsnm{Davis},~\bfnm{M.~A.~S.}\binits{M.~A.~S.}}
(\byear{2012}).
\bhowpublished{NHC data archive. {National Hurricane Center}. {Available at \url{http://1.usa.gov/cCcwTg}}.}
\end{bmisc}
\bptok{imsref}%
\endbibitem

\bibitem[\protect\citeauthoryear{Kass and Raftery}{1994}]{kass:raftery:1994}
\begin{barticle}[author]
\bauthor{\bsnm{Kass},~\bfnm{R.~E.}\binits{R.~E.}} \AND
\bauthor{\bsnm{Raftery},~\bfnm{A.~E.}\binits{A.~E.}}
(\byear{1994}).
\btitle{Bayes factors}.
\bjournal{J. Amer. Statist. Assoc.}
\bvolume{90}
\bpages{773--795}.
\end{barticle}
\bptok{imsref}%
\endbibitem

\bibitem[\protect\citeauthoryear{Kratz and Resnick}{1996}]{kratz:resnick:1996}
\begin{barticle}[mr]
\bauthor{\bsnm{Kratz},~\bfnm{Marie}\binits{M.}} \AND
\bauthor{\bsnm{Resnick},~\bfnm{Sidney~I.}\binits{S.~I.}}
(\byear{1996}).
\btitle{The {$\mathrm{QQ}$}-estimator and heavy tails}.
\bjournal{Comm. Statist. Stochastic Models}
\bvolume{12}
\bpages{699--724}.
\bid{doi={10.1080/15326349608807407}, issn={0882-0287}, mr={1410853}}
\end{barticle}
\bptok{imsref}%
\endbibitem

\bibitem[\protect\citeauthoryear{McLachlan and Jones}{1988}]{mclachlan:jones:1988}
\begin{barticle}[author]
\bauthor{\bsnm{McLachlan},~\bfnm{G.~J.}\binits{G.~J.}} \AND
\bauthor{\bsnm{Jones},~\bfnm{P.~N.}\binits{P.~N.}}
(\byear{1988}).
\btitle{Fitting mixture models to grouped and truncated data via the EM algorithm}.
\bjournal{Biometrics}
\bvolume{44}
\bpages{571--578}.
\end{barticle}
\bptok{imsref}%
\endbibitem

\bibitem[\protect\citeauthoryear{Mitzenmacher}{2004}]{mitzenmacher04}
\begin{barticle}[mr]
\bauthor{\bsnm{Mitzenmacher},~\bfnm{Michael}\binits{M.}}
(\byear{2004}).
\btitle{A brief history of generative models for power law and lognormal distributions}.
\bjournal{Internet Math.}
\bvolume{1}
\bpages{226--251}.
\bid{issn={1542-7951}, mr={2077227}}
\end{barticle}
\bptok{imsref}%
\endbibitem

\bibitem[\protect\citeauthoryear{Mitzenmacher}{2006}]{mitzenmacher06}
\begin{barticle}[author]
\bauthor{\bsnm{Mitzenmacher},~\bfnm{M.}\binits{M.}}
(\byear{2006}).
\btitle{The future of power law research}.
\bjournal{Internet Math.}
\bvolume{2}
\bpages{525--534}.
\end{barticle}
\bptok{imsref}%
\endbibitem

\bibitem[\protect\citeauthoryear{Newman}{2005}]{newman:2005}
\begin{barticle}[author]
\bauthor{\bsnm{Newman},~\bfnm{M.~E.~J.}\binits{M.~E.~J.}}
(\byear{2005}).
\btitle{Power laws, {Pareto} distributions and {Z}ipf's law}.
\bjournal{Contemporary Physics}
\bvolume{46}
\bpages{323--351}.
\end{barticle}
\bptok{imsref}%
\endbibitem

\bibitem[\protect\citeauthoryear{Noether}{1963}]{noether:1963}
\begin{barticle}[mr]
\bauthor{\bsnm{Noether},~\bfnm{G.~E.}\binits{G.~E.}}
(\byear{1963}).
\btitle{Note on the {K}olmogorov statistic in the discrete case}.
\bjournal{Metrika}
\bvolume{7}
\bpages{115--116}.
\bid{issn={0026-1335}, mr={0158462}}
\end{barticle}
\bptok{imsref}%
\endbibitem

\bibitem[\protect\citeauthoryear{{Orphanet Report Series, Rare Diseases collection}}{2011}]{prevalence11}
\begin{bmisc}[author]
\borganization{Orphanet Report Series, Rare Diseases collection}
(\byear{2011}).
\bhowpublished{Prevalence of rare diseases: Bibliographic data.
Available at \url{http://bit.ly/MezSZ6}.}
\end{bmisc}
\bptok{imsref}%
\endbibitem

\bibitem[\protect\citeauthoryear{Persing and Montgomery}{2003}]{persing:montgomery:2003}
\begin{barticle}[author]
\bauthor{\bsnm{Persing},~\bfnm{J.}\binits{J.}} \AND
\bauthor{\bsnm{Montgomery},~\bfnm{M.~T.}\binits{M.~T.}}
(\byear{2003}).
\btitle{Hurricane superintensity}.
\bjournal{J. Atmospheric Sci.}
\bvolume{60}
\bpages{2349--2371}.
\end{barticle}
\bptok{imsref}%
\endbibitem

\bibitem[\protect\citeauthoryear{Press et~al.}{1992}]{press:etal:1992}
\begin{bbook}[mr]
\bauthor{\bsnm{Press},~\bfnm{William~H.}\binits{W.~H.}},
\bauthor{\bsnm{Teukolsky},~\bfnm{Saul~A.}\binits{S.~A.}},
\bauthor{\bsnm{Vetterling},~\bfnm{William~T.}\binits{W.~T.}} \AND
\bauthor{\bsnm{Flannery},~\bfnm{Brian~P.}\binits{B.~P.}}
(\byear{1992}).
\btitle{Numerical Recipes in {C}: The Art of Scientific Computing},
\bedition{2nd} ed.
\bpublisher{Cambridge Univ. Press},
\blocation{Cambridge}.
\bid{mr={1201159}}
\end{bbook}
\bptok{imsref}%
\endbibitem

\bibitem[\protect\citeauthoryear{Rao}{1947}]{rao46}
\begin{barticle}[mr]
\bauthor{\bsnm{Rao},~\bfnm{C.~Radhakrishna}\binits{C.~R.}}
(\byear{1947}).
\btitle{Minimum variance and the estimation of several parameters}.
\bjournal{Proc. Cambridge Philos. Soc.}
\bvolume{43}
\bpages{280--283}.
\bid{mr={0019904}}
\bptnote{check year}%
\end{barticle}
\bptok{imsref}%
\endbibitem

\bibitem[\protect\citeauthoryear{Rao}{1957}]{rao:1957}
\begin{barticle}[mr]
\bauthor{\bsnm{Rao},~\bfnm{C.~Radhakrishna}\binits{C.~R.}}
(\byear{1957}).
\btitle{Maximum likelihood estimation for the multinomial distribution}.
\bjournal{Sankhy\=a}
\bvolume{18}
\bpages{139--148}.
\bid{issn={0972-7671}, mr={0105183}}
\end{barticle}
\bptok{imsref}%
\endbibitem

\bibitem[\protect\citeauthoryear{Reed and Hughes}{2002}]{reed:hughes:2002}
\begin{barticle}[author]
\bauthor{\bsnm{Reed},~\bfnm{W.~J.}\binits{W.~J.}} \AND
\bauthor{\bsnm{Hughes},~\bfnm{B.~D.}\binits{B.~D.}}
(\byear{2002}).
\btitle{From gene families and genera to income and {internet} file sizes: Why power laws are so common in nature}.
\bjournal{Phys. Rev. E (3)}
\bvolume{66}
\bpages{067103}.
\end{barticle}
\bptok{imsref}%
\endbibitem

\bibitem[\protect\citeauthoryear{Reiss and Thomas}{2007}]{reiss:thomas:2007}
\begin{bbook}[mr]
\bauthor{\bsnm{Reiss},~\bfnm{R.-D.}\binits{R.-D.}} \AND
\bauthor{\bsnm{Thomas},~\bfnm{M.}\binits{M.}}
(\byear{2007}).
\btitle{Statistical Analysis of Extreme Values with Applications to Insurance, Finance, Hydrology and Other Fields},
\bedition{3rd} ed.
\bpublisher{Birkh\"auser},
\blocation{Basel}.
\bid{mr={2334035}}
\end{bbook}
\bptok{imsref}%
\endbibitem

\bibitem[\protect\citeauthoryear{Richardson}{1960}]{richardson:1960}
\begin{bbook}[author]
\bauthor{\bsnm{Richardson},~\bfnm{Lewis~F.}\binits{L.~F.}}
(\byear{1960}).
\btitle{Statistics of Deadly Quarrels}.
\bpublisher{The Boxwood Press},
\blocation{Pittsburgh}.
\end{bbook}
\bptok{imsref}%
\endbibitem

\bibitem[\protect\citeauthoryear{Schultze and Steinebach}{1996}]{schultze:steinebach:1996}
\begin{barticle}[mr]
\bauthor{\bsnm{Schultze},~\bfnm{J.}\binits{J.}} \AND
\bauthor{\bsnm{Steinebach},~\bfnm{J.}\binits{J.}}
(\byear{1996}).
\btitle{On least squares estimates of an exponential tail coefficient}.
\bjournal{Statist. Decisions}
\bvolume{14}
\bpages{353--372}.
\bid{issn={0721-2631}, mr={1437826}}
\end{barticle}
\bptok{imsref}%
\endbibitem

\bibitem[\protect\citeauthoryear{Shinokazi et~al.}{1964}]{shinokazi:etal:1964}
\begin{barticle}[author]
\bauthor{\bsnm{Shinokazi},~\bfnm{K.}\binits{K.}},
\bauthor{\bsnm{Yoda},~\bfnm{K.}\binits{K.}},
\bauthor{\bsnm{Hozumi},~\bfnm{K.}\binits{K.}} \AND
\bauthor{\bsnm{Kira},~\bfnm{T.}\binits{T.}}
(\byear{1964}).
\btitle{A quantitative analysis of plant form---{The pipe model theory} {II}: Further evidence of the theory and its application in forest ecology}.
\bjournal{Japanese Journal of Ecology}
\bvolume{14}
\bpages{133--139}.
\end{barticle}
\bptok{imsref}%
\endbibitem

\bibitem[\protect\citeauthoryear{Sornette}{2006}]{sornette:2006}
\begin{bbook}[mr]
\bauthor{\bsnm{Sornette},~\bfnm{Didier}\binits{D.}}
(\byear{2006}).
\btitle{Critical Phenomena in Natural Sciences: Chaos, Fractals, Selforganization and Disorder: Concepts and Tools},
\bedition{2nd} ed.
\bpublisher{Springer},
\blocation{Berlin}.
\bid{mr={2220576}}
\end{bbook}
\bptok{imsref}%
\endbibitem

\bibitem[\protect\citeauthoryear{Stoev, Michailidis and Taqqu}{2011}]{stoev:michailidis:taqqu:2006}
\begin{barticle}[mr]
\bauthor{\bsnm{Stoev},~\bfnm{Stilian~A.}\binits{S.~A.}},
\bauthor{\bsnm{Michailidis},~\bfnm{George}\binits{G.}} \AND
\bauthor{\bsnm{Taqqu},~\bfnm{Murad~S.}\binits{M.~S.}}
(\byear{2011}).
\btitle{Estimating heavy-tail exponents through max self-similarity}.
\bjournal{IEEE Trans. Inform. Theory}
\bvolume{57}
\bpages{1615--1636}.
\bid{doi={10.1109/TIT.2010.2103751}, issn={0018-9448}, mr={2815838}}
\end{barticle}
\bptok{imsref}%
\endbibitem

\bibitem[\protect\citeauthoryear{Stone}{1974}]{stone:1974}
\begin{barticle}[mr]
\bauthor{\bsnm{Stone},~\bfnm{M.}\binits{M.}}
(\byear{1974}).
\btitle{Cross-validatory choice and assessment of statistical predictions}.
\bjournal{J. R. Stat. Soc. Ser. B Stat. Methodol.}
\bvolume{36}
\bpages{111--147}.
\bid{issn={0035-9246}, mr={0356377}}
\bptnote{check related}%
\end{barticle}
\bptok{imsref}%
\endbibitem

\bibitem[\protect\citeauthoryear{{Storm Prediction Center}}{2011}]{spc11}
\begin{bmisc}[author]
\borganization{Storm Prediction Center}
(\byear{2011}).
\bhowpublished{Severe weather database files (1950--2011).
Available at\break \url{http://1.usa.gov/Lj7cC9}}.
\end{bmisc}
\bptok{imsref}%
\endbibitem

\bibitem[\protect\citeauthoryear{Stumpf and Porter}{2012}]{stumpf:porter:2012}
\begin{barticle}[mr]
\bauthor{\bsnm{Stumpf},~\bfnm{Michael~P.~H.}\binits{M.~P.~H.}} \AND
\bauthor{\bsnm{Porter},~\bfnm{Mason~A.}\binits{M.~A.}}
(\byear{2012}).
\btitle{Critical truths about power laws}.
\bjournal{Science}
\bvolume{335}
\bpages{665--666}.
\bid{doi={10.1126/science.1216142}, issn={0036-8075}, mr={2932329}}
\end{barticle}
\bptok{imsref}%
\endbibitem

\bibitem[\protect\citeauthoryear{Tate and Hye}{1973}]{tate:hye:1973}
\begin{barticle}[author]
\bauthor{\bsnm{Tate},~\bfnm{M.~W.}\binits{M.~W.}} \AND
\bauthor{\bsnm{Hye},~\bfnm{L.~A.}\binits{L.~A.}}
(\byear{1973}).
\btitle{Inaccuracy of the $\chi^2$ test of goodness of fit when expected frequencies are small}.
\bjournal{J. Amer. Statist. Assoc.}
\bvolume{68}
\bpages{836--841}.
\end{barticle}
\bptok{imsref}%
\endbibitem

\bibitem[\protect\citeauthoryear{Virkar and Clauset}{2014}]{suppl}
\begin{bmisc}[auto]
\bauthor{\bsnm{Virkar},~\bfnm{Yogesh}\binits{Y.}} \AND
\bauthor{\bsnm{Clauset},~\bfnm{Aaron}\binits{A.}}
(\byear{2014}).
\bhowpublished{Supplement to ``Power-law distributions in binned empirical data.'' DOI:\doiurl{10.1214/13-AOAS710SUPP}.}
\end{bmisc}
\bptok{imsref}%
\endbibitem

\bibitem[\protect\citeauthoryear{Vuong}{1989}]{vuong:1989}
\begin{barticle}[mr]
\bauthor{\bsnm{Vuong},~\bfnm{Quang~H.}\binits{Q.~H.}}
(\byear{1989}).
\btitle{Likelihood ratio tests for model selection and nonnested hypotheses}.
\bjournal{Econometrica}
\bvolume{57}
\bpages{307--333}.
\bid{doi={10.2307/1912557}, issn={0012-9682}, mr={0996939}}
\end{barticle}
\bptok{imsref}%
\endbibitem

\bibitem[\protect\citeauthoryear{Wasserman}{2004}]{wasserman:2003}
\begin{bbook}[mr]
\bauthor{\bsnm{Wasserman},~\bfnm{Larry}\binits{L.}}
(\byear{2004}).
\btitle{All of Statistics: A Concise Course in Statistical Inference}.
\bpublisher{Springer},
\blocation{New York}.
\bid{mr={2055670}}
\bptnote{check year}%
\end{bbook}
\bptok{imsref}%
\endbibitem

\bibitem[\protect\citeauthoryear{West, Enquist and Brown}{2009}]{west09}
\begin{barticle}[author]
\bauthor{\bsnm{West},~\bfnm{G.~B.}\binits{G.~B.}},
\bauthor{\bsnm{Enquist},~\bfnm{B.~J.}\binits{B.~J.}} \AND
\bauthor{\bsnm{Brown},~\bfnm{J.~H.}\binits{J.~H.}}
(\byear{2009}).
\btitle{A general quantitative theory of forest structure and dynamics}.
\bjournal{Proc. Natl. Acad. Sci. USA}
\bvolume{106}
\bpages{7040--7045}.
\end{barticle}
\bptok{imsref}%
\endbibitem

\bibitem[\protect\citeauthoryear{World Glacier Monitoring Service and
National Snow and Ice Data Center}{2012}]{wgi12}
\begin{bmisc}[author]
\borganization{World Glacier Monitoring Service}
\AND
\borganization{National Snow and Ice Data Center}
(\byear{2012}).
\bhowpublished{World glacier inventory.
Available at \url{http://bit.ly/MhLdt6}.}
\end{bmisc}
\bptok{imsref}%
\endbibitem

\bibitem[\protect\citeauthoryear{Yamamoto and Kobayashi}{1993}]{yamamoto93}
\begin{barticle}[author]
\bauthor{\bsnm{Yamamoto},~\bfnm{K.}\binits{K.}} \AND
\bauthor{\bsnm{Kobayashi},~\bfnm{S.}\binits{S.}}
(\byear{1993}).
\btitle{Analysis of crown structure based on the pipe model theory}.
\bjournal{Journal of the Japanese Forestry Society}
\bvolume{75}
\bpages{445--448}.
\end{barticle}
\bptok{imsref}%
\endbibitem

\end{thebibliography}
\end{document}